\documentclass[twocolumn,preprint]{aastex631}

\usepackage{hyperref}
\usepackage{xcolor}
\usepackage{enumitem}

\newcommand{\msun}{M_\odot}

\newcommand{\HeI}{\hbox{{\rm He}~{\sc i}}}

\newcommand{\NII}{\hbox{{\rm [N}~{\sc ii}{\rm ]}}}

\newcommand{\Ha}{\hbox{{\rm H}$\alpha$}}

\defcitealias{Reines_Volonteri_2015}{RV15}

\shorttitle{The Skyfire Survey}
\shortauthors{Kocevski et al.}
\graphicspath{{./}{figures/}}

\begin{document}

\title{\large \bf Skyfire: A Spectroscopic Census of Little Red Dots and Broad-Line AGN \\in the CEERS Field}

\suppressAffiliations

\author[0000-0002-8360-3880]{Dale D. Kocevski}
\affiliation{Department of Physics and Astronomy, Colby College, Waterville, ME 04901, USA}

\author[0000-0003-1282-7454]{Anthony J. Taylor}
\affiliation{Department of Astronomy, The University of Texas at Austin, Austin, TX, USA}
\affiliation{Cosmic Frontier Center, The University of Texas at Austin, Austin, TX, USA}

\author[0000-0003-2984-6803]{Masafusa Onoue}
\affiliation{Kavli Institute for the Physics and Mathematics of the Universe (Kavli IPMU, WPI), The University of Tokyo, Chiba 277-8583, Japan}

\author[0000-0001-9840-4959]{Kohei Inayoshi}
\affiliation{Kavli Institute for Astronomy and Astrophysics, Peking University, Beijing 100871, China}

\author[0000-0001-8519-1130]{Steven L. Finkelstein}
\affiliation{Department of Astronomy, The University of Texas at Austin, Austin, TX, USA}
\affiliation{Cosmic Frontier Center, The University of Texas at Austin, Austin, TX, USA}

\author[0000-0002-0786-7307]{Guillermo Barro}
\affiliation{Department of Physics, University of the Pacific, Stockton, CA 90340 USA}

\author[0009-0002-6578-8110]{Jingsong Guo}
\affiliation{Department of Astronomy, School of Physics, Peking University, Beijing 100871, China}

\author[0000-0002-7959-8783]{Pablo Arrabal Haro}
\altaffiliation{NASA Postdoctoral Fellow}
\affiliation{Astrophysics Science Division, NASA Goddard Space Flight Center, 8800 Greenbelt Rd, Greenbelt, MD 20771, USA}

\author[0000-0002-1410-0470]{Jonathan R. Trump}
\affiliation{Department of Physics, 196A Auditorium Road, Unit 3046, University of Connecticut, Storrs, CT 06269, USA}
\email{jonathan.trump@uconn.edu}

\author[0000-0003-2366-8858]{Rebecca L. Larson}
\altaffiliation{Giacconi Postdoctoral Fellow}
\affiliation{Space Telescope Science Institute, 3700 San Martin Drive, Baltimore, MD 21218, USA}

\author[0000-0001-5414-5131]{Mark Dickinson}
\affiliation{NSF's National Optical-Infrared Astronomy Research Laboratory, 950 N. Cherry Ave., Tucson, AZ 85719, USA}

\author[0000-0001-7503-8482]{Casey Papovich}
\affiliation{Department of Physics and Astronomy, Texas A\&M University, College Station, TX, 77843-4242 USA}
\affiliation{George P.\ and Cynthia Woods Mitchell Institute for Fundamental Physics and Astronomy, Texas A\&M University, College Station, TX, 77843-4242 USA}

\author[0000-0001-9879-7780]{Fabio Pacucci}
\affiliation{Center for Astrophysics $\vert$ Harvard \& Smithsonian, Cambridge, MA 02138, USA}
\affiliation{Black Hole Initiative, Harvard University, Cambridge, MA 02138, USA}

\author[0000-0003-4528-5639]{Pablo G. P\'erez-Gonz\'alez}
\affiliation{Centro de Astrobiolog\'{\i}a (CAB), CSIC-INTA, Ctra. de Ajalvir km 4, Torrej\'on de Ardoz, E-28850, Madrid, Spain}

\author[0000-0002-3301-3321]{Michaela Hirschmann}
\affiliation{Institute of Physics, Laboratory of Galaxy Evolution, Ecole Polytechnique Fédérale de Lausanne (EPFL), Observatoire de Sauverny, 1290 Versoix, Switzerland}

\author[0000-0001-8688-2443]{Elizabeth J.\ McGrath}
\affiliation{Department of Physics and Astronomy, Colby College, Waterville, ME 04901, USA}

\author[0000-0002-8360-3880]{Brenda L. Jones}
\affiliation{Department of Physics and Astronomy, University of Maine, Orono, ME 04469, USA}


\author[0000-0001-7151-009X]{Nikko J. Cleri}
\affiliation{Department of Astronomy and Astrophysics, The Pennsylvania State University, University Park, PA 16802, USA}
\affiliation{Institute for Computational and Data Sciences, The Pennsylvania State University, University Park, PA 16802, USA}
\affiliation{Institute for Gravitation and the Cosmos, The Pennsylvania State University, University Park, PA 16802, USA}

\author[0000-0001-8047-8351]{Kelcey Davis}
\altaffiliation{NSF Graduate Research Fellow}
\affiliation{Department of Physics, 196A Auditorium Road, Unit 3046, University of Connecticut, Storrs, CT 06269, USA}
\affil{Los Alamos National Laboratory, Los Alamos, NM 87545, USA}

\author[0000-0002-7831-8751]{{Mauro} {Giavalisco}}
\affiliation{University of Massachusetts Amherst, 710 North Pleasant Street, Amherst, MA 01003-9305, USA}

\author[0000-0001-9440-8872]{Norman A. Grogin}
\affiliation{Space Telescope Science Institute, Baltimore, MD, USA}

\author[0000-0001-6251-4988]{Taylor A. Hutchison}
\altaffiliation{NASA Postdoctoral Fellow}
\affiliation{Astrophysics Science Division, NASA Goddard Space Flight Center, 8800 Greenbelt Rd, Greenbelt, MD 20771, USA}

\author[0000-0001-9187-3605]{Jeyhan S. Kartaltepe}
\affiliation{Laboratory for Multiwavelength Astrophysics, School of Physics and Astronomy, Rochester Institute of Technology, 84 Lomb Memorial Drive, Rochester, NY 14623, USA}

\author[0000-0002-5537-8110]{Allison Kirkpatrick}
\affiliation{Department of Physics and Astronomy, University of Kansas, Lawrence, KS 66045, USA}

\author[0000-0002-9393-6507]{Gene C. K. Leung}
\affiliation{MIT Kavli Institute for Astrophysics and Space Research, 77 Massachusetts Avenue, Cambridge, MA 02139, USA}

\author[0000-0002-6748-6821]{Rachel S.~Somerville}
\affiliation{Center for Computational Astrophysics, Flatiron Institute, 162 5th Avenue, New York, NY 10010, USA}

\author[0000-0003-3903-6935]{Stephen M.~Wilkins} %
\affiliation{Astronomy Centre, University of Sussex, Falmer, Brighton BN1 9QH, UK}
\affiliation{Institute of Space Sciences and Astronomy, University of Malta, Msida MSD 2080, Malta}

\affiliation{}

\begin{abstract}

We present the Skyfire program, a 21-hour Cycle 3 JWST/NIRSpec survey with the G395M medium-resolution grating covering five pointings in the Extended Groth Strip.  The survey is designed to carry out a systematic census of faint, broad-line AGN candidates with a range of rest-optical colors identified photometrically at $z>3$ by the Cosmic Evolution Early Release Science (CEERS) Survey.  Our primary targets include photometrically-selected Little Red Dots (LRDs), blue extreme emission line galaxies (EELGs), and X-ray-detected AGN.  We present spectroscopic redshifts for 178 sources observed by Skyfire, as well as a catalog of 34 sources with broad emission lines in the redshift range $2.7 < z < 6.5$.  Our broad-line sample includes 18 LRDs, which brings the spectroscopic completeness of LRDs with $\beta_{\rm opt}>-0.02$ in the CEERS field to 73\%.  We explore the prevalence of broad emission lines in photometrically-selected LRDs as a function of their rest-frame continuum slope and observed color distributions.  We find the broad-line detection fraction in LRDs remains high at relatively blue rest-optical colors and extends smoothly into the bluer regime occupied by Little Blue Dots (LBDs).  We discuss the implications of this finding for LRD-LBD unification scenarios.  We also find that only 18\% (3/17) of EELGs selected primarily for their high-equivalent-width emission lines and compact morphologies exhibit broad emission lines, suggesting these criteria alone are poor predictors of broad-line activity.  We present a revised set of LRD selection criteria that captures bluer sources by extending down to $\beta_{\rm opt}=-0.52$.  Using this new threshold, we find that $80.9^{+4.6}_{-7.5}\%$ of photometrically-selected LRDs brighter than 26.5 in F444W show broad emission lines and that LRDs make up 54\% of the overall broad-line population identified in the CEERS field.

\end{abstract}


\section{Introduction}

It is now well established that supermassive black holes (SMBHs) reside at the centers of nearly all massive galaxies and that the growth of these black holes (BHs) is intimately connected to the evolution of their hosts (e.g., \citealt{Hopkins_2008, DiMatteo_2008, KH_2013, Madau_Dickinson_2014}).  As a result, SMBHs and the active galactic nuclei (AGN) that they power have become key components in most galaxy evolution models (e.g., \citealt{somerville08, dubois16, zhu20, dave19}).  However, several open issues remain in our understanding of their formation and early growth, including the nature of the BH seed population and how the BH-galaxy scaling relationships observed in the local universe \citep{Kormendy1995,Magorrian98, Gebhardt00, Ferrarese00} are established and maintained. Addressing these questions requires characterizing the growth of SMBHs from their earliest seeds through the peak of cosmic activity, and constraining the distribution of BH masses and their accretion rates in the first billion years of cosmic history \citep{Volonteri_2010, Inayoshi_2020, Habouzit22}.

Prior to the launch of the James Webb Space Telescope \citep[JWST;][]{Gardner23}, our understanding of BH growth in the early universe was largely shaped by wide-field quasar surveys carried out at optical and near-infrared wavelengths \citep{fan01, willott10, jiang16, Matsuoka2016, Yang23}. These surveys revealed hundreds of quasars at $z > 5$ powered by BHs with masses of order $10^{9}$ M$_{\odot}$, but remain sensitive to only the most luminous and extreme systems.  JWST has fundamentally changed this picture by enabling the detection of a large population of faint broad-line AGN at $z > 3$  \citep{Kocevski23b, Uebler23, Harikane23, Maiolino24, Greene24, Matthee24, Taylor25, Hviding25}.  These sources are orders of magnitude less luminous and more common than bright quasars and are powered by the least massive SMBHs known in the early universe \citep{onoue23, Taylor25, Jones26}. Because they are more representative of the underlying BH population than luminous quasars, these faint AGN offer a promising route to constraining models of BH seeding \citep{Pacucci22, Li24} and the early coevolution of SMBHs with their host galaxies \citep{Habouzit22, Inayoshi_2022, Pacucci_2023_overmassive}.

A substantial fraction of these broad-line AGN appear heavily reddened in the rest-frame optical with relatively blue colors in the rest-frame ultraviolet \citep{Kocevski23b, Furtak23, Greene24, Matthee24, Kocevski25}. Compact sources with this ``v-shaped" spectral energy distribution have come to be known as little red dots \citep[LRDs;][]{Matthee24} and represent one of the most unexpected discoveries of the JWST era.  Photometric searches have found LRDs to be remarkably ubiquitous at high redshifts, reaching number densities of order $10^{-5}$ Mpc$^{-3}$ at $z\sim5-6$ \citep{Barro24, Kokorev24, Labbe25} before declining rapidly below $z\sim4$ (\citealt{Kocevski25, Ma25, Zhuang_2025, Weibel26}; although see \citealt{Bisigello25}).  

Spectroscopic follow-up confirms that 70–90\% of photometrically selected LRDs exhibit broad Balmer emission lines \citep{Greene24, Kocevski25, Hviding25}.  While the precise nature of LRDs remains a topic of intense debate \citep[e.g.,][]{Naidu25, Madau26, Pacucci26, Pablo26}, the detection of extreme Balmer breaks in some LRDs with strengths that exceed those achievable with evolved stellar populations \citep{Naidu25, deGraaff25, Taylor25b} suggests that they are likely powered by a previously unrecognized population of gas-enshrouded, accreting SMBHs \citep{Inayoshi_Maiolino25, Kokorev26, deGraaff26}.

\begin{figure*}[t]
\centering
\includegraphics[width=\linewidth]{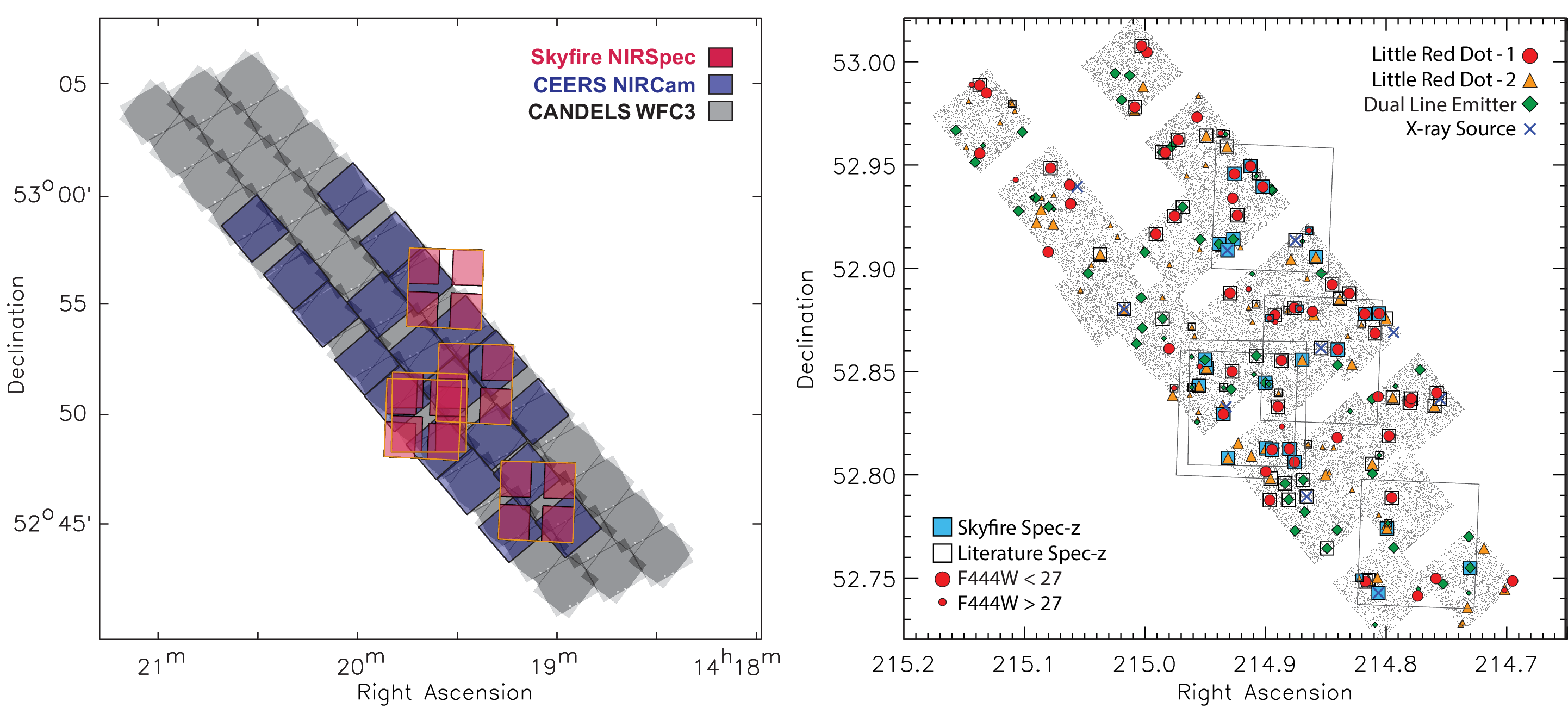}
\caption{(\emph{left}) The pointing centers of the five NIRSpec MSA masks observed by Skyfire relative to the JWST/NIRCam imaging from CEERS (blue) and HST/WFC3 imaging from CANDELS (grey). (\emph{right}) The distribution of high priority targets in the CEERS field. LRD-1 sources are drawn from the LRD sample of \citet{Kocevski25}, while LRD-2 sources are drawn from the LRD samples of \citet{Kokorev24} and \cite{Barro24}.  Dual line emitters are blue, broad-line AGN candidates with strong emission lines from the catalog of \citet{Guo25}.  Sources observed by Skyfire are shown with filled blue squares, while those with prior spectroscopic observations are highlighted with open squares. Larger/smaller symbols are used for sources brighter/fainter than 27th magnitude in F444W. \label{fig:layout}}
\end{figure*}

Recently, \citet{Barro26} reported broad-line detections among LRDs with bluer rest-optical colors than the red colors cuts used by canonical LRD selection criteria \citep{Labbe23, Kocevski25}, indicating that LRDs constitute a more diverse population than initially recognized, extending beyond the most extreme red objects.  The relationship between LRDs and their bluer, broad-line emitting counterparts, recently dubbed ``little blue dots" \citep[LBDs,][]{Asada26, Brazzini26}, remains unsettled.  Recent studies have argued that LRDs and LBDs may not be physically distinct populations, with viewing angle and line-of-sight gas density plausibly responsible for turning an intrinsically blue, compact broad-line AGN into a red one \citep{Billand26, Madau26}.  A key difficulty in establishing a unification scenario is that while LRDs have generally been targeted for spectroscopic follow-up due to their distinctive colors, compact AGN candidates with bluer rest-frame optical continua more closely resemble ordinary star-forming galaxies and have typically only turned up serendipitously within larger spectroscopic campaigns \citep[e.g.,][]{Taylor25, Hviding25}.  This implies that an unknown fraction of the broad-line AGN population has been missed by red-color-based searches. Resolving this will require spectroscopic samples that span the full range of rest-frame optical colors among compact, high-redshift AGN candidates, rather than samples preselected for their red colors alone.

In this paper, we present the Skyfire program (GO-5718; PIs Kocevski \& Guo), a Cycle 3 JWST/NIRSpec multi-object spectroscopic survey with the G395M medium-resolution ($R\sim1000$) grating in the Extended Groth Strip (EGS) field.  The survey is designed to carry out the first systematic census of faint broad-line AGN candidates with a range of rest-optical colors identified photometrically by the Cosmic Evolution Early Release Science (CEERS) Survey \citep{bagley22, Finkelstein_2025}. The motivation for the survey is to increase the spectroscopic completeness of LRDs, LBDs, and extreme emission line galaxies in the CEERS field in order to complete the census of faint, broad-line emitting AGN in the early Universe.  

In Section 2, we present the Skyfire survey design and target selection, while Section 3 describes the NIRSpec observations and our data reduction. Section 4 discusses our methodology to measure spectroscopic redshifts and broad-line widths. Section 5 presents our spectroscopic redshift catalog and broad emission line sample, while Section 6 presents a handful of individual sources of interest.  Throughout this paper we use the following cosmological parameters: $H_{0} = 70~{\rm km~s^{-1}~Mpc^{-1}; \Omega_{tot}, \Omega_{\Lambda}, \Omega_{m} = 1, 0.7, 0.3}$.

\section{Survey Design}

Skyfire is designed to perform a spectroscopic census of faint AGN through the detection of their broad emission lines.  The survey utilizes NIRSpec observations with the G395M/F290LP grating/filter pair, which provides optimal spectral coverage ($3-5\mu$m) and resolution ($R\sim1000$) needed for this purpose.  The spectral coverage allows both the H$\alpha$ line and the H$\beta$+[O{\sc iii}] multiplet to be detected at $z\sim5$ and the H$\beta$ line out to $z\sim9.4$, while the velocity resolution ($\Delta v\sim300$ km s$^{-1}$) is sufficient to resolve broad emission (${\rm FWHM}\sim1000$ km s$^{-1}$) originating near the AGN accretion disk from a narrow component from the host galaxy.  It also allows for the deblending of the H$\beta$+[O{\sc iii}] complex and the detection of broad components in the individual [O{\sc iii}] lines.  The lack of broad emission in these forbidden lines can help rule out galactic outflows as the source of any broad emission observed in the Balmer lines.  

The EGS was chosen as the target field for Skyfire due to the wealth of multi-wavelength data available in the field.  This includes 1-4 $\mu$m NIRCam imaging from CEERS \citep{bagley22, Finkelstein25}, 5–21$\mu$m MIRI observations from both CEERS \citep{YangG23} and the MIRI EGS Galaxy and Active Galactic Nucleus (MEGA) survey \citep{Backhaus25}, 0.6-1.6$\mu$m Hubble Space Telescope (HST) ACS and WFC3 imaging from the Cosmic Assembly Near-infrared Deep Extragalactic Legacy Survey \citep[CANDELS;][]{grogin11, koekemoer11}, and deep (800 ksec) Chandra X-ray observations \citep{nandra15}.  The field also has extensive optical and infrared spectroscopy from the All-Wavelength Extended Groth Strip International Survey \citep[AEGIS;][]{Davis07}, 3D-HST \citep{brammer12}, CEERS \citep{Pirzkal17,ArrabalHaro23}, the CANDELS-Area Prism Epoch of Reionization Survey \citep[CAPERS;][]{Donnan25}, the Red Unknowns: Bright Infrared Extragalactic Survey \citep[RUBIES;][]{deGraaff25}, The High-(Redshift+Ionization) Line Search \citep[THRILS;][]{Hutchison25}, C3PO  (Papovich et al., in prep), and the MEGA NIRSpec survey (Kirkpatrick et al., in prep).

\begin{table}
\caption{Skyfire Observations}\vspace{0mm}
\begin{center}
\vspace{-0.2in}
\begin{tabular}{ccccc}
\hline
\hline
Visit & R.A.    & Dec.    & APA   & Obs Date  \\
      & (J2000) & (J2000) & (deg) &           \\
 \hline
1:1  &  14:19:34  &  52:55:35  &  267.67  &  2025-06-19   \\
1:2  &  14:19:25  &  52:51:16  &  267.63  &  2025-06-21   \\
1:3  &  14:19:41  &  52:49:42  &  267.69  &  2025-06-23   \\
1:4  &  14:19:22  &  52:48:49  &  267.58  &  2025-06-24   \\
1:5  &  14:19:39  &  52:49:59  &   89.20  &  2025-12-17   \\
\hline 
\end{tabular}
\label{tab_pointings}
\end{center}
\end{table}

\subsection{Target Selection}

The primary targets for Skyfire are LRDs and blue, broad-line AGN candidates.  The LRD sample is split into two subsamples.  The first subsample is the LRD catalog from \cite{Kocevski25}, which we refer to hereafter as the LRD-1 sample.  This sample consists of sources selected based on having a ``v-shaped'' spectral energy distribution (SED) with a strongly rising rest-optical continuum and a compact morphology.  In a recent analysis by \citet{Hviding25}, it was reported that while this sample has a high accuracy in identifying LRDs with broad-line emission, it becomes increasingly incomplete towards bluer rest-optical colors.  As a result, we supplement this sample with the LRD catalogs of \citet{Kokorev24} and \cite{Barro24}, which use multi-color selection that picks up LRDs with bluer rest-optical colors.  \citet{Hviding25} find that while the completeness of this secondary subsample is higher than that of \citet{Kocevski25}, it suffers from a decreased accuracy, presumably due to contamination from galaxies with strong emission lines or dusty star-forming galaxies at lower redshifts.  We refer to this subsample as the LRD-2 sample hereafter.  

Our sample of blue, broad-line AGN candidates is drawn from the catalog of dual line emitters (DLE) from \citet{Guo25}.  These sources are extreme emission line galaxies (EELGs) selected photometrically to have strong H$\beta$+[O{\sc iii}] and H$\alpha$ lines. The sample has a mean rest-frame equivalent width for H$\beta$ and H$\alpha$ of 990\AA~and 1050\AA, respectively.   The selection technique of \citet{Guo25} was successfully used by \citet{onoue23} to discover a z=5.24 AGN, CEERS 2782, which was later confirmed to have broad emission lines by \citet{Kocevski23b}. 
We refer to this subsample as the DLE sample hereafter.  Finally, we added X-ray sources at $z>2.5$ drawn from the AEGIS-X survey \citep{nandra15} to our primary target sample.

We cross-matched our primary targets with a spectroscopic redshift compilation from the Dawn JWST
Archive\footnote{https://dawn-cph.github.io/dja} \citep[DJA; ][]{deGraaff24, Heintz24}. Sources previously observed with NIRSpec's M or H gratings that resulted in spectra where the width of the Balmer lines could be measured were moved from our primary sample to our filler sample (described below).  Sources observed with the lower-resolution NIRSpec PRISM were kept in the primary sample and their redshifts updated if the source previously lacked a spectroscopic redshift.

For our filler catalog, we employed a multiwavelength catalog described in detail in \citet{Hutchison25}.  In short, a combined source catalog was constructed using public CEERS NIRCam and MIRI imaging and the CANDELS EGS catalogs of \citet{Stefanon17} and \citet{Finkelstein22}. For the CEERS data, photometry and photometric redshifts were drawn from the UNICORN photometric catalog (see \S3.1; S. Finkelstein et al., in prep.).  For each source, near-infrared magnitudes were drawn preferentially from JWST/NIRCam where available or from HST/WFC3 or Spitzer/IRAC otherwise. Photometric redshifts were drawn from three catalogs in priority order — the CEERS UNICORN NIRCam-based redshifts, then the CANDELS HST+Spitzer values, then the CANDELS EGS catalog of \citet{Finkelstein22}. Spectroscopic redshifts followed the same tiered priority scheme, drawing first from the CEERS NIRSpec catalog, then the DJA, then CANDELS, and finally ground-based DEIMOS \citep{Newman13} and Keck/MOSFIRE \citep{Sanders21, Larson22, Jung18, Stawinski24} programs.

\begin{figure*}[t]
\centering
\includegraphics[width=0.95\linewidth]{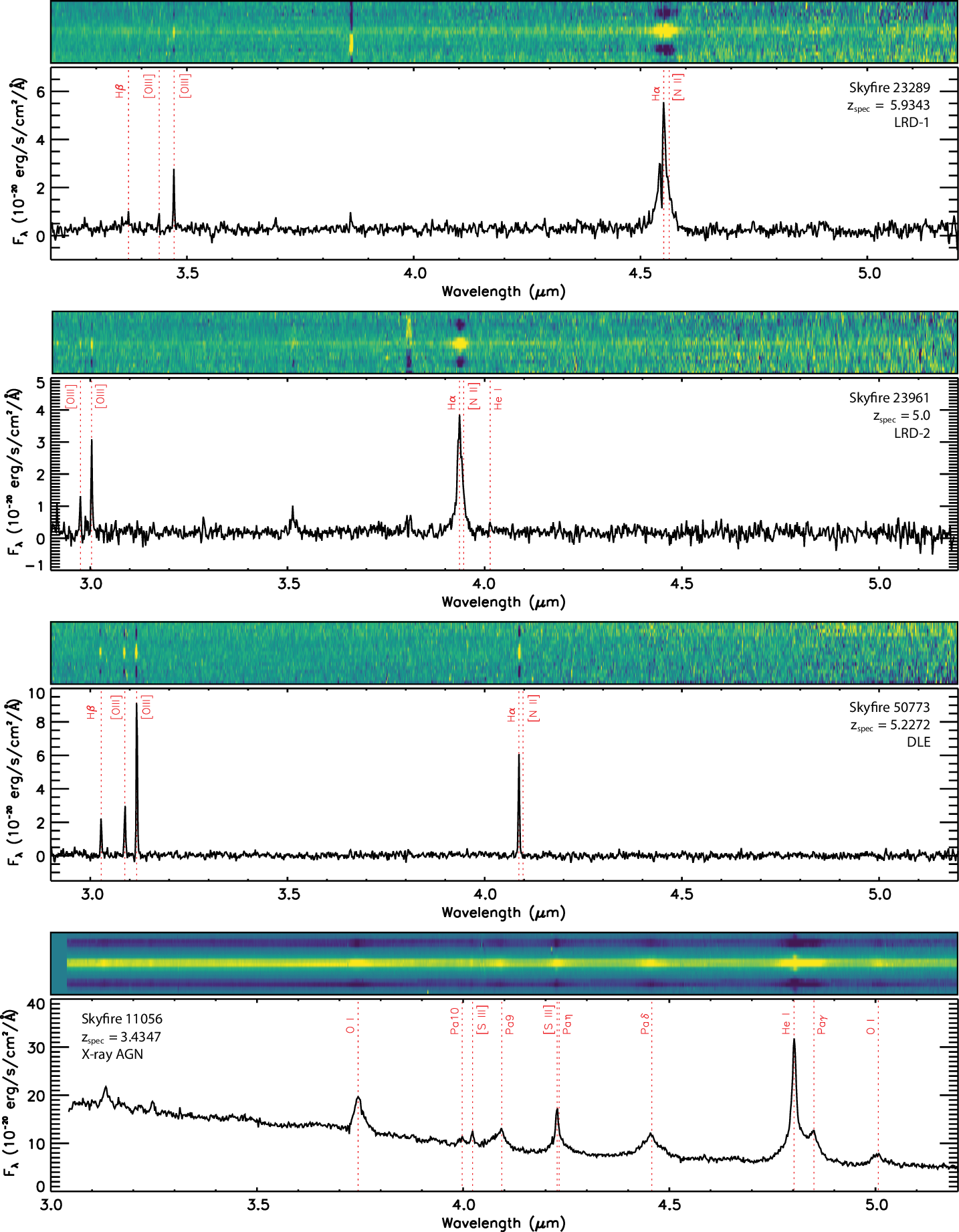}
\caption{Example Skyfire NIRSpec spectra taken in the G395M grating.  The sources shown are examples of each of the four high-priority target classes observed by Skyfire: LRD-1, LRD-2, DLEs, and X-ray AGN. The locations of several prominent emission lines are noted.
\label{fig:spectra}}
\end{figure*}

\begin{figure*}[t]
\centering
\includegraphics[width=6in]{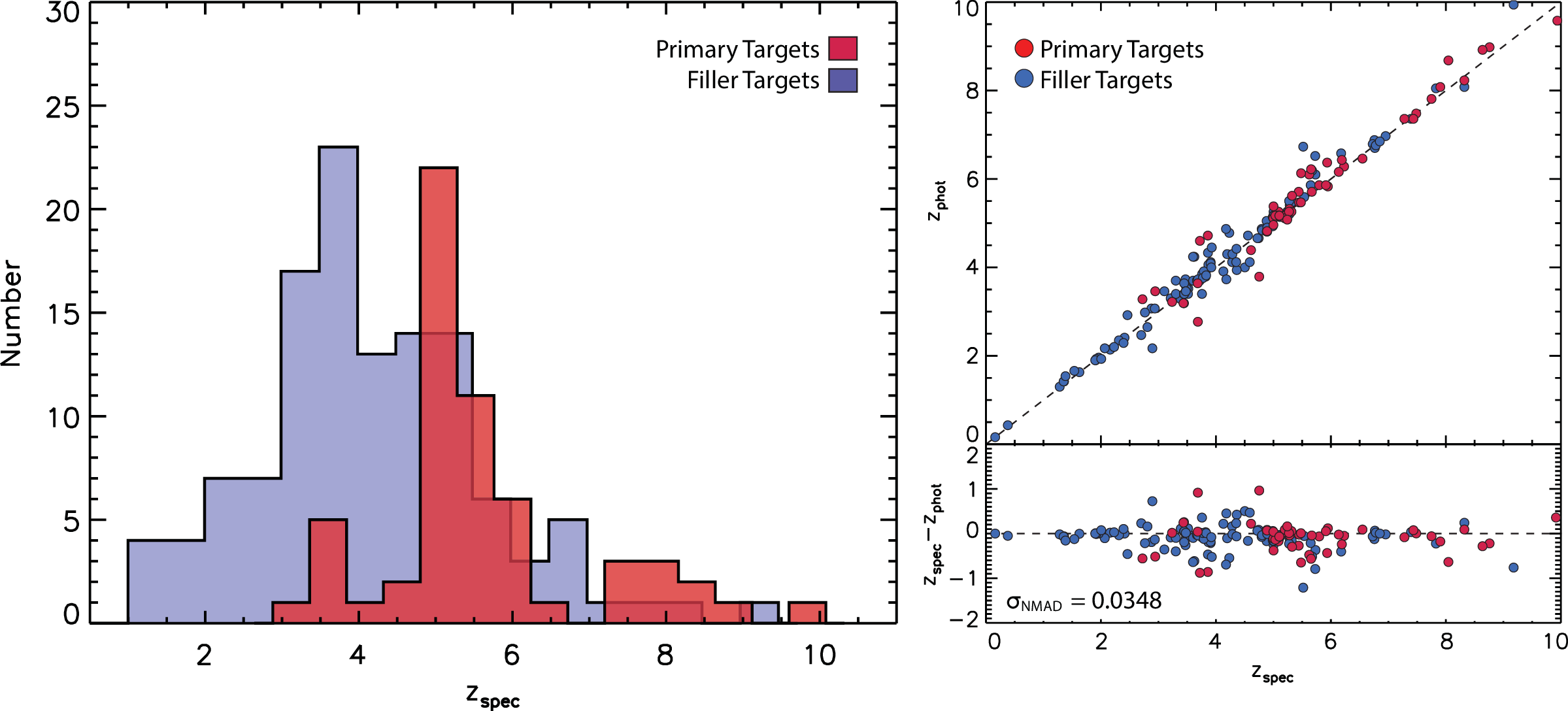}
\caption{(\emph{left}) The spectroscopic redshift distribution of the primary (red) and filler (blue) sources observed by Skyfire. (\emph{right}) A comparison of the photometric and spectroscopic redshifts of the sources observed by Skyfire.  We find good agreement between our measured spectroscopic redshifts and our prior photometric redshifts, measuring a normalized median absolute deviation of $\sigma_{\rm NMAD} = 0.0348$.   \label{fig:redshifts}}
\end{figure*}

MSA optimization was done in two stages.  First we ran the eMPT software suite \citep{Bonaventura23} on our primary target list to identify pointing centers that would maximize the number of high-priority candidate AGN on slits given our assigned aperture position angle (APA). Primary targets with ${\rm F444W} < 27$ were given uniform weighting as priority class 1 objects in eMPT, while fainter sources were assigned priority class 2.  With candidate pointing centers computed, we then used the MSA Planning Tool (MPT) in APT to optimize the masks.  We used a weighting scheme that downweights sources at lower redshift, as well as fainter sources that might not be detected.  The formula used can be expressed as:
\begin{equation} \label{eq_weights}
W = \sqrt{F_{\rm nJy} * z^6}.
\end{equation} 
where $W$ is the MSA weight, $F_{\rm nJy}$ is the flux density in nanojanskies of the source in the F444W filter or Spitzer Ch2 when outside of the JWST coverage, and $z$ is the best available redshift for each source.  The weights were normalized to range from 1 to 1000.  Sources in our primary target sample were manually boosted by an additional factor of 1000.  Optimal masks were defined as those that achieved the highest combined weight for all targets allocated MSA shutters.  Through this process, we identified six pointing centers that maximized the number of high-priority candidate AGN on slits and the combined weights of our filler targets.

\section{Observations \& Data Reduction} \label{sec:obs_data} 

The initial round of Skyfire observations were carried out on 2025 June 19-24.  We used 3-shutter slitlets and a 3-shutter nodding pattern for each pointing with the NRSIRS2 readout pattern.  We used 16 groups per integration and 2 integrations per exposure, resulting in a total exposure time of 7090 sec.  Six pointings were attempted at an APA of $267^{\circ}$, but two suffered MSA target acquisition failures.  One of these pointings was re-executed on 2025 Dec 17 at an APA of $89^{\circ}$.  The pointing centers of our final five MSA masks are shown in Figure \ref{fig:layout} and listed in Table \ref{tab_pointings}.  The number of our primary targets observed in each mask ranges from 11 to 15, while the total number of targets observed ranges from 54 to 61.  The location of the primary targets observed in each pointing is shown on the right side of Figure \ref{fig:layout}.  

The Skyfire data was reduced as described in \cite{Taylor25} using version~1.20.2 of the JWST Science Calibration Pipeline with the Calibration Reference Data System (CRDS) mapping 1464, starting from the Level~1 uncalibrated data products (``\_uncal.fits'' files) available on MAST. 

The reduced two-dimensional (2D) spectra (``s2d'') have a rectified trace with straight spatial and wavelength axes. To optimize the extraction of one-dimensional (1D) spectra from the 2D spectra, we perform a weighted extraction based on the methodology of \cite{Horne86}. Briefly, for a given spectrum, we take the median of the 2D spectrum along the spectral direction to produce a spatial profile for the source. We then identify the central peak of this profile, which corresponds to the source's spectral trace. We then set all pixels in the spatial profile that are not a part of this central feature to zero and normalize the area under this masked spatial profile to one. We then use the normalized profile as the variable \textbf{P} in Table~1 of \cite{Horne86} and follow the prescription given therein to extract an optimized 1D spectrum. Examples of our reduced 1D and 2D spectra, for each high-priority target category (LRD-1, LRD-2, DLEs, and X-ray AGN), are shown in Figure \ref{fig:spectra}. 

\begin{table*}
\caption{The Skyfire Survey Spectroscopic Redshift Catalog}
\begin{center}
\begin{tabular}{cccccccc}
\hline
\hline
ID & RA & Dec & $z_{\rm spec}$ &  $z_{\rm qual}$  & $z_{\rm phot}$  & F444W & Source \\
   & (J2000) & (J2000) &  &  &  &  & Flag  \\
\hline
 299  &  214.805753  &  52.878045  &  5.6229  &  4  &  6.10  &  26.12  &  1 \\
 467  &  214.903179  &  52.946261  &  5.1027  &  4  &  5.17  &  26.18  &  3 \\
1090  &  214.912510  &  52.949436  &  6.2290  &  4  &  6.28  &  26.80  &  1 \\
1867  &  214.817678  &  52.877856  &  5.4457  &  4  &  5.47  &  25.55  &  1 \\
2086  &  214.858256  &  52.905539  &  5.4806  &  4  &  6.13  &  24.31  &  2 \\
2847  &  214.875185  &  52.913482  &  2.9434  &  4  &  3.46  &  23.57  &  4 \\
3023  &  214.853818  &  52.897569  &  5.2931  &  4  &  5.26  &  26.76  &  3 \\
3519  &  214.925759  &  52.945659  &  5.0885  &  4  &  5.26  &  26.01  &  1 \\
4979  &  214.879011  &  52.904279  &  5.6548  &  4  &  6.22  &  26.92  &  2 \\
5632  &  214.932394  &  52.938252  &  4.8883  &  4  &  4.82  &  24.49  &  5 \\
\hline
\end{tabular}
\label{tab_redshift1}
\end{center}
\vspace{-0.15in}
\tablecomments{Source Flag: 1 = LRD-1, 2 = LRD-2, 3 = DLE, 4 = X-ray AGN, 5 = Filler.  This table is available in its entirety at https://github.com/dalekocevski/Kocevski26}
\end{table*}

\subsection{NIRCam Imaging \& Photometric Redshifts}  

In this study, we make use of {\it JWST} NIRCam imaging from the Cosmic Evolution Early Release Science Survey (CEERS; \citet{Finkelstein25}). The NIRCam data were processed using the {\it JWST} Calibration Pipeline\footnote{\url{http://jwst-pipeline.readthedocs.io/en/latest/}} (versions 1.8.5 and 1.10.2, respectively) with custom modifications described in \citet{Finkelstein25} and \citet{bagley22}. The resulting images were registered to the same World Coordinate System reference frame (based on Gaia DR1.2; \citealt{Gaia16}) and combined into a single mosaic for each field using the drizzle algorithm with an inverse variance map weighting \citep{fruchter02,Casertano00} via the Resample step in the pipeline.  The final mosaics in all fields have pixel scales of 0\farcs03/pixel. 

Source detection and photometry on the NIRCam mosaics were computed on PSF-matched images using \texttt{SExtractor} \citep{bertin96} version~2.25.0 in two-image mode, with an inverse-variance weighted combination of the PSF-matched F277W and F356W images as the detection image.  Photometry was measured in all of the available NIRCam bands in each field, as well as the F606W and F814W HST bands using public data from the CANDELS surveys \citep{grogin11, koekemoer11}.  Two runs of \texttt{SExtractor} were conducted in a hot+cold setup similar to the process described in \citet{Galametz13}.  
Sources are first selected with conservative (cold) detection parameters, designed not to split up large, bright galaxies, followed by a more aggressive (hot) run, which is performed
to identify fainter objects.  Objects from the hot catalog that fall outside the segmentation map from the cold run are added to the final photometry catalog.  Our analysis of the NIRCam imaging data is similar to that described in \citet{Finkelstein24}; hence, we refer the reader there for additional details.

Photometric redshifts were computed using the \textsc{eazy} \citep{brammer08} software package for all sources in our photometric catalogs.  \textsc{eazy} fits non-negative linear combinations of user-supplied templates to derive probability distribution functions (PDFs) for the redshift, based on the quality of fit of the various template combinations to the observed photometry for a given source.  We use the default template set ``tweak\_fsps\_QSF\_12\_v3", which consists of 12 templates derived from the stellar population synthesis code FSPS \citep{Conroy10}, as well as the six bluer templates created by \cite{Larson22}, as described in \cite{Finkelstein23}.  A flat redshift prior with respect to luminosity was assumed and redshifts from $z = 0$ to 20 were considered.  We also perform a “low-redshift” run with the maximum redshift set to $z = 7$ to allow for visualization of the best-fitting low-redshift model for sources found to be at $z>9$.

\section{Redshift and Line Measurements} \label{sec:methodology} 

An initial measurement of spectroscopic redshifts was carried out using an automated emission line fitting algorithm. 
The software uses the cataloged photometric redshift of each source to identify emission lines and fit their centroids.  The spectra of all sources were then manually inspected to check the automated redshift calculation, measure redshifts for sources where the algorithm failed to compute a redshift, and grade the quality of the redshift measurement.  We assigned quality flags based on the strength (i.e., signal-to-noise ratio, S/N) and number of the identified emission lines and the agreement with existing photometric redshift estimates.  The quality scheme for the derived redshifts is as follows:
   
\vspace{0.1in}
\hangindent=0.25in \hangafter=1 $\bullet$ 4.0: Multiple high S/N ($\gtrsim 3$) emission lines 
  
\hangindent=0.25in \hangafter=1 $\bullet$ 3.0: Multiple low S/N ($\lesssim 3$) emission lines. 
  
\hangindent=0.25in \hangafter=1 $\bullet$ 2.0: Single emission line + $z_{spec}$ agrees with $z_{phot}$

\hangindent=0.25in \hangafter=1 $\bullet$ 1.0: Unidentified features
  
\hangindent=0.25in \hangafter=1 $\bullet$ 0.0: No features
  
\vspace{0.1in}

\noindent Sources with quality $\leq 1$ have their redshifts set to NaN in our final redshift catalog (see \S\ref{sec:zcat}).  

In addition to redshift measurements, we also perform a search for broad emission lines across the entire target sample (primary and filler).  We measure line fluxes and widths in the G395M/F290LP spectra using a Levenberg-Marquardt least-squares method implemented by the \texttt{mpfit} IDL code\footnote{https://pages.physics.wisc.edu/$\sim$craigm/idl/fitting.html} as described in \citet{Kocevski23b}.  We simultaneously fit multiple Gaussians for features in the Balmer and Paschen line regions, depending on the redshift of the source.  To account for broad components, we fit lines with two Gaussians: one narrow with width $<350$~km~s$^{-1}$ and one broad with width $>350$~km~s$^{-1}$.  The line centers, widths, and fluxes are all free parameters and the broad and narrow components can be kinematically offset from each other.  When fitting \Ha\, we also include additional Gaussian components for the \NII~$\lambda\lambda$6550, 6585 doublet. While the \Ha\ and \NII~$\lambda$6585 lines are separated by roughly three times the resolution limit, the lines will blend together in the presence of a sufficiently broad \Ha\ component.  We account for this by constraining the line widths and relative line centers of the \NII\ doublet to that of narrow \Ha.  If a source shows signs of blue-shifted absorption in their Balmer lines similar to the ones reported in \citet{Matthee_2023} and \citet{Kocevski25}, we add an additional absorption component to our model.  

Uncertainties on our line measurements are derived using a Monte Carlo approach.  For each spectral element, we perform 1000 random draws from a Gaussian distribution whose mean is set to the measured flux at that wavelength and whose standard deviation is set to the error in that measurement.  Our line fitting is then repeated on all 1000 mock spectra, and standard deviations are calculated as half the 16th-to-84th percentile range of the resulting distributions. 

To test the statistical significance of the broad line components, we used the approach described in \citet{Taylor25}.  We fit each source with a model that includes and excludes the broad component and compute the Bayesian Information Criterion (BIC) for each model fit.  We next use the difference in the BIC between the models to determine which model fits best for each candidate line (with the lower BIC model preferred).  We require a $\Delta {\rm BIC} > 6$ in favor of the broad-line model over the narrow line model and a $S/N>5$ for the broad component, suggesting that the inclusion of the broad-line significantly improves the overall spectral fit without overfitting or adding unnecessary components to the model.

\section{Redshift Catalog} \label{sec:zcat} 

In this section, we provide the redshift catalog for the Skyfire target sample.  Skyfire observed 292 unique sources over 5 masks.  Of these, 41 are LRDs (24 from the LRD-1 sample and 17 from the LRD-2 sample), 26 are DLEs, and 4 are X-ray AGN.  We successfully measured spectroscopic redshifts (${\rm quality} > 1$) for 178 sources.  Another 30 sources have unidentified features (${\rm quality} = 1$) and 84 sources lacked features from which to base a redshift measurement (${\rm quality}=0$).  The redshift distribution of our primary and filler samples is shown in Figure \ref{fig:redshifts}.

\begin{figure*}[t]
\centering
\includegraphics[width=\linewidth]{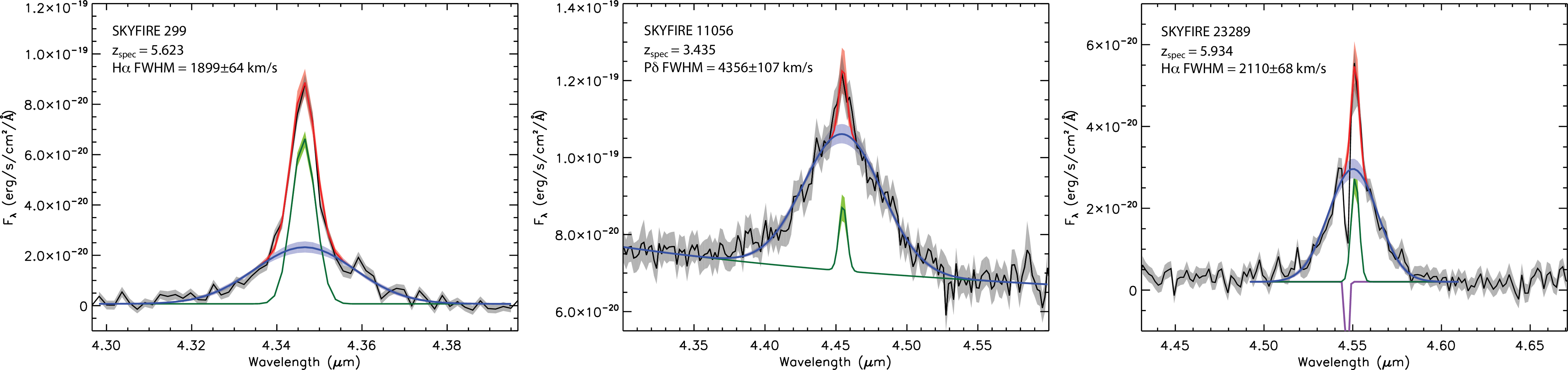}
\caption{Example NIRSpec spectra with uncertainties (grey shaded region) taken in the G395M grating of three sources that feature broad emission lines. Green lines show the best-fit Gaussian for the narrow emission line component, blue lines show the best-fit broad component, and red lines show the best overall (narrow plus broad) fit to the emission line.  The FWHM of the broad component (corrected for instrument broadening) is shown in the upper left of each panel.\label{fig:BLfits_examples}}
\end{figure*}

Our spectroscopic redshifts allow us to assess the accuracy of the photometric redshifts of our target sample.  A comparison between the photometric and spectroscopic redshifts is shown on the right panel of Figure \ref{fig:redshifts}. We can quantify the agreement between the two using the normalized median absolute deviation ($\sigma_{\rm NMAD}$; \citealt{Ilbert09}), defined as $1.48\times {\rm median}(|\Delta z|/(1+z_{\rm spec}))$. We find that our measured redshifts are in overall good agreement with our photometric redshift estimates, with $\sigma_{\rm NMAD}=0.0348$, which is typical of robust photometric redshifts based solely on broadband photometry \citep[e.g., ][]{Dahlen13}.  We note that $\sigma_{\rm NMAD}$ for the primary sample is lower than that for the filler sample: 0.0165 versus 0.0397, respectively.  This is largely driven by the DLE sample, which has a $\sigma_{\rm NMAD}$ of 0.0138.  These sources are selected photometrically due to boosted line emission affecting the F277W and F410M filters, which preferentially identifies sources in a narrow redshift range near $z\sim5$, resulting in a relatively low scatter.  The LRD sample (LRD-1 plus LRD-2) has a scatter comparable to the filler sample, with $\sigma_{\rm NMAD}$ of 0.0281.  This suggests that the usual SEDs of LRDs do not result in an increased photometric redshift uncertainty as is commonly found with other forms of AGN \citep[e.g., ][]{Salvato09}.

\begin{table*}
\caption{Catalog of Skyfire Sources with Broad Emission Lines}
\begin{center}
\begin{tabular}{cccccccc}
\hline
\hline
ID & R.A. & Dec & $z_{\rm spec}$ &  FWHM  &  $F_{\rm broad}$  & $\log\,(M_{\rm BH}/{\rm \msun})$ & Line \\
   & (J2000) & (J2000) & & (km s$^{-1}$) &  (10$^{-18}$ erg s$^{-1}$ cm$^{-2}$)  &  &  \\
\hline
       299  &  214.805753  &  52.878045  &  5.6229  &  $ 1896\pm  63 $  &  $  6.63\pm 0.15 $  & $7.31^{+0.03}_{-0.03}$  & $ {\rm H}\alpha  $     \\
      1090  &  214.912510  &  52.949436  &  6.2290  &  $ 1141\pm  94 $  &  $  2.83\pm 0.16 $  & $6.73^{+0.08}_{-0.09}$  & $ {\rm H}\alpha  $     \\
      1867  &  214.817678  &  52.877856  &  5.4457  &  $ 1640\pm 185 $  &  $  2.23\pm 0.14 $  & $6.94^{+0.11}_{-0.12}$  & $ {\rm H}\alpha  $     \\
      2086  &  214.858256  &  52.905539  &  5.4806  &  $ 2111\pm  97 $  &  $  7.70\pm 0.28 $  & $7.42^{+0.05}_{-0.05}$  & $ {\rm H}\alpha  $     \\
      2847  &  214.875185  &  52.913482  &  2.9434  &  $ 1247\pm 157 $  &  $  2.93\pm 0.21 $  & -                       & $ \HeI~\lambda10830$  \\
      3519  &  214.925759  &  52.945659  &  5.0885  &  $ 1946\pm 126 $  &  $  3.84\pm 0.15 $  & $7.17^{+0.06}_{-0.07}$  & $ {\rm H}\alpha  $     \\
      7761  &  214.840034  &  52.860648  &  3.8597  &  $ 1985\pm  50 $  &  $ 13.41\pm 0.20 $  & $7.31^{+0.03}_{-0.03}$  & $ {\rm H}\alpha  $     \\
      9382  &  214.926772  &  52.914039  &  5.0730  &  $  596\pm  79 $  &  $  1.30\pm 0.22 $  & $5.89^{+0.14}_{-0.17}$  & $ {\rm H}\alpha  $     \\
     11056  &  214.931614  &  52.908693  &  3.4347  &  $ 4356\pm 107 $  &  $ 24.66\pm 0.42 $  & -                       & $ {\rm Pa}\delta $     \\   
     12630  &  214.869654  &  52.855608  &  5.1033  &  $ 2215\pm 443 $  &  $  1.35\pm 0.17 $  & $7.08^{+0.19}_{-0.23}$ & $ {\rm H}\alpha  $     \\ 
     17484  &  214.799323  &  52.774004  &  5.0334  &  $ 1524\pm 186 $  &  $  1.67\pm 0.10 $  & $6.78^{+0.11}_{-0.13}$ & $ {\rm H}\alpha  $     \\
     20952  &  214.880185  &  52.812566  &  5.2802  &  $ 1882\pm 123 $  &  $  4.42\pm 0.20 $  & $7.19^{+0.07}_{-0.07}$ & $ {\rm H}\alpha  $     \\
     21749  &  214.876034  &  52.806109  &  5.7958  &  $ 1601\pm  70 $  &  $  8.20\pm 0.19 $  & $7.22^{+0.04}_{-0.04}$ & $ {\rm H}\alpha  $     \\
     22814  &  214.943538  &  52.849030  &  5.2413  &  $ 1299\pm 269 $  &  $  2.70\pm 0.42 $  & $6.75^{+0.20}_{-0.24}$ & $ {\rm H}\alpha  $     \\
     23082  &  214.949011  &  52.851744  &  3.7210  &  $ 3948\pm 168 $  &  $ 21.90\pm 1.24 $  & $8.01^{+0.05}_{-0.05}$ & $ {\rm H}\alpha  $     \\
     23289  &  214.894556  &  52.812162  &  5.9343  &  $ 2109\pm  68 $  &  $  9.69\pm 0.18 $  & $7.51^{+0.03}_{-0.03}$ & $ {\rm H}\alpha  $     \\
     23961  &  214.899699  &  52.812840  &  5.0000  &  $ 1702\pm  92 $  &  $  5.14\pm 0.18 $  & $7.11^{+0.05}_{-0.06}$ & $ {\rm H}\alpha  $     \\
     24689  &  214.806408  &  52.742786  &  3.6858  &  $ 1766\pm  74 $  &  $  6.74\pm 0.36 $  & $7.05^{+0.05}_{-0.05}$ & $ {\rm H}\alpha  $     \\
     25941  &  214.934979  &  52.829364  &  5.9469  &  $ 1424\pm 329 $  &  $  1.73\pm 0.14 $  & $6.81^{+0.20}_{-0.25}$ & $ {\rm H}\alpha  $     \\
     26079  &  214.955188  &  52.843018  &  5.0015  &  $ 2182\pm  84 $  &  $ 12.45\pm 0.32 $  & $7.51^{+0.04}_{-0.04}$ & $ {\rm H}\alpha  $     \\
     30148  &  214.931119  &  52.808158  &  6.5491  &  $ 1671\pm 119 $  &  $  2.96\pm 0.17 $  & $7.10^{+0.07}_{-0.08}$ & $ {\rm H}\alpha  $     \\
     44039  &  214.822229  &  52.750196  &  5.9128  &  $ 1641\pm 271 $  &  $  1.76\pm 0.13 $  & $6.93^{+0.15}_{-0.18}$ & $ {\rm H}\alpha  $     \\
     83652  &  214.871873  &  52.880432  &  5.6669  &  $ 1202\pm 128 $  &  $  1.50\pm 0.11 $  & $6.60^{+0.10}_{-0.12}$ & $ {\rm H}\alpha  $     \\
     92141  &  214.896761  &  52.875796  &  6.1365  &  $ 1897\pm 217 $  &  $  1.13\pm 0.15 $  & $6.99^{+0.12}_{-0.14}$ & $ {\rm H}\alpha  $     \\
    504101  &  214.947772  &  52.836333  &  4.8814  &  $ 2437\pm 192 $  &  $  2.49\pm 0.15 $  & $7.27^{+0.08}_{-0.09}$ & $ {\rm H}\alpha  $     \\ 
    504403  &  214.809743  &  52.739684  &  2.7007  &  $ 1103\pm 199 $  &  $  1.48\pm 0.18 $  & -                      & $ \HeI~\lambda10830$  \\ 
    513817  &  214.884590  &  52.844341  &  3.6240  &  $ 1441\pm 119 $  &  $  2.45\pm 0.32 $  & $6.65^{+0.10}_{-0.11}$ & $ {\rm H}\alpha  $     \\
    514153  &  214.746291  &  52.747477  &  3.2128  &  $ 2601\pm 273 $  &  $  3.24\pm 0.24 $  & -                      & $ {\rm Pa}\delta$     \\  
    530722  &  214.871720  &  52.914307  &  4.3569  &  $  574\pm  67 $  &  $  3.09\pm 1.09 $  & $5.96^{+0.16}_{-0.20}$ & $ {\rm H}\alpha  $     \\
    534080  &  214.949994  &  52.855774  &  5.2761  &  $ 1388\pm 271 $  &  $  0.72\pm 0.15 $  & $6.55^{+0.20}_{-0.24}$ & $ {\rm H}\alpha  $     \\
    542588  &  214.857161  &  52.903543  &  3.6015  &  $  979\pm 165 $  &  $  2.68\pm 0.68 $  & $6.32^{+0.19}_{-0.22}$     &  ${\rm H}\alpha $  \\  
    547783  &  214.755946  &  52.759043  &  3.6886  &  $ 1676\pm 153 $  &  $  1.20\pm 0.12 $  & $6.65^{+0.10}_{-0.11}$ & $ {\rm H}\alpha  $     \\
    549030  &  214.883004  &  52.830524  &  3.7898  &  $  400\pm  50 $  &  $  1.17\pm 0.20 $  & $5.37^{+0.14}_{-0.16}$ & $ {\rm H}\alpha  $     \\
\hline
\end{tabular}
\label{tab_blagn}
\end{center}
\vspace{-0.15in}
\end{table*}

We present our spectroscopic redshift catalog in Table \ref{tab_redshift1}. The details of the table columns are given below:

\vspace{0.1in}  
\indent    1. Skyfire ID \\*
\indent    2. Right ascension (J2000)\\*
\indent    3. Declination (J2000)\\*
\indent    4. Spectroscopic Redshift \\*
\indent    5. Spectroscopic Redshift quality flag (see \S4)\\*
\indent    6. Photometric Redshift\\*
\indent    7. Magnitude (AB; F444W or Spitzer Ch2)\\*
\indent    8. Source Category Flag (see below)\\*

\noindent For the source category flag, we assign the following flags to each type of source observed: flag 1 are LRD-1, flag 2 are LRD-2, flag 3 are DLEs, flag 4 are X-ray sources, and flag 10 are filler targets.

\section{Broad-Line AGN Sample} \label{sec:BL_AGN} 

One of the primary science goals of Skyfire is the detection and characterization of faint, broad-line AGN.  In this section, we present a sample of broad-line emitters identified with Skyfire using the selection criteria outlined in \S4.  We find a total of 34 sources with broad emission lines in the redshift range $2.7<z<6.5$.  Of these, 33 are new detections, while one source, Skyfire-549755, is a reobservation of RUBIES-EGS-28812 reported in \cite{Kocevski25}.  Example G395M spectra of three sources with broad-line emission, in regions near their broad components, is shown in Figure \ref{fig:BLfits_examples}, while the spectra of all 33 sources are shown in Figure \ref{fig:BLfits1} in Appendix A. Information about our broad-line sample, including our best-fit line widths and fluxes are listed in Table \ref{tab_blagn}

The vast majority of the broad lines detected are H$\alpha$ (29), while the others include the \HeI $\lambda10830$ (3) and Pa$\gamma$ (2) lines. The broad-line widths that we measure for these lines range from 500 to 4000 km s$^{-1}$.  Two sources in our sample, Skyfire-23082 and Skyfire-23289, also show absorption in their Balmer lines, a feature that has become increasingly common among LRDs \citep{Matthee_2023, Kocevski25, Davis26, Matthee26}.

\begin{figure*}[t]
\centering
\includegraphics[width=\linewidth]{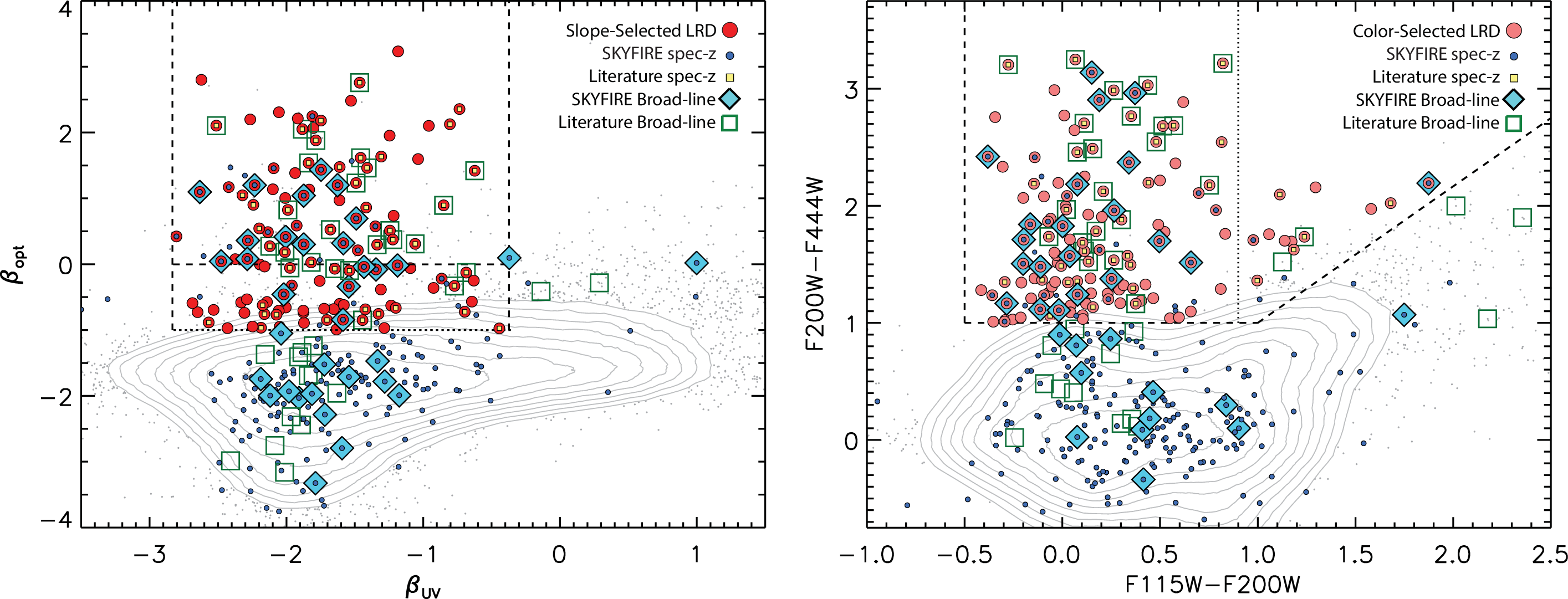}
\caption{(\emph{left}) The rest-frame optical and UV spectral slopes, $\beta_{\rm opt}$ and $\beta_{\rm UV}$ measured in the CEERS field for galaxies at $z>2$ detected in the F444W filter with a ${\rm SNR}>12$.  The horizontal and vertical dashed lines denote the LRD selection criteria of \citet{Kocevski25}, while the dotted line is a lower rest-optical threshold of $\beta_{\rm opt} = -0.52$. The red circles are sources that satisfy the slope-based LRD selection criteria of \citet{Kocevski25} with this lower $\beta_{\rm opt}$ cut.  Sources with spectroscopic redshifts from Skyfire and the literature are shown as small blue circles and yellow squares, respectively.  Sources with broad emission lines from Skyfire and the literature are highlighted with filled blue diamonds and unfilled green squares, respectively. (\emph{right}) The F115W-F200W versus F200W-F444W color distribution of the same galaxies shown in the left panel.  The light red circles are sources that satisfy the color-based LRD selection criteria of \citet{Barro26}.  The dashed lines denote that study's selection wedge.  We find that the broad-line detection fraction remains high below the canonical cuts used to identify LRDs and likely extends smoothly into the bluer regime occupied by LBDs, suggesting these sources may be a continuous distribution of broad-line emitters and not be two distinct classes of objects\label{fig:slope-color}}
\end{figure*}

Among the broad-line sources, 18 are LRDs (11 from LRD-1 and 7 from LRD-2), 3 are DLEs, 3 are X-ray sources, and 10 are filler targets.  Of the 7 LRD-2 sources, we find that 4 would satisfy the LRD selection criteria of \citet{Kocevski25} if a lower rest-optical slope cut is adopted ($\beta_{\rm opt}=-0.52$ instead of -0.02, which translates to F444W-F277W colors of 0.75 and 1.0, respectively).  The remaining three sources were flagged by \citet{Kocevski25} as potential extreme emission line galaxy (EELG) contaminants and therefore not included in their sample.

Of the 19 DLEs targeted for spectroscopic follow-up, redshifts were successfully measured for 17, of which only 3 exhibit broad emission lines.  These sources are compact EELGs drawn from the sample of \citet{Guo25} and have a mean H$\alpha$ rest-frame equivalent width of 1050\AA.  This suggests that selecting sources based primarily on the presence of high-equivalent-width emission lines and compact morphologies is not particularly effective at isolating broad-line AGN and that samples selected in this manner are at risk of being contaminated by star-forming galaxies.  That said, we note that 11 of our 19 DLE targets are relatively faint, with ${\rm F444W} < 27$, resulting in low signal-to-noise spectra that may have limited our ability to detect broad components even if they were present.  A firmer assessment of this method's robustness will require deeper spectroscopy of the fainter subsample to rule out sensitivity-driven incompleteness as the cause of the low broad-line detection rate.  This result reiterates the difficulty of identifying blue broad-line AGN given that their colors make them hard to distinguish from the general galaxy population.

\section{Black Hole Mass Estimates} \label{sec:BH_masses} 

We estimate the BH masses of our broad-line sample assuming that their broad emission lines trace the kinematics of gas that is gravitationally bound to and in virial orbits around the central BH.  We follow the methodology of \citet{Jones26} and make use of the single-epoch scaling relationship presented in \citet{reines13}:  
 \begin{equation} \label{eq: GH05}
 M_{\rm BH} = 3.7 \times 10^6 \left( \frac{L_{\rm H\alpha}}{10^{42}\ {\rm erg\ s^{-1}}}\right)^{0.47}  \left(\frac{{\rm FWHM_{\rm H\alpha}}}{10^3\ {\rm km\ s^{-1}}} \right)^{2.06} M_\odot.
 \end{equation} 
Here $L_{\rm H\alpha}$ and ${\rm FWHM_{\rm H\alpha}}$ are the luminosity and FWHM of the broad H$\alpha$ line.  Our measured BH masses for all sources where broad H$\alpha$ is detected are listed in Table \ref{tab_blagn}.  Formal errors on our mass estimates are derived using the MCMC approach described in \S4, but the \citet{reines13} relationship has an uncertainty of roughly 0.5 dex \citep{Reines_Volonteri_2015} which will dominate over our measurement uncertainties.

We note several important caveats regarding our reported BH masses.  First, in calculating our BH masses, we do not correct our line luminosities for dust attenuation.  The level of dust obscuration in LRDs remains uncertain and there is a growing consensus that their red colors are the result of being embedded in a dense gas cocoon as opposed to dust extinction \citep{Inayoshi_Maiolino25, Naidu25, deGraaff25, Taylor25b, Kokorev26}.   Given that LRDs comprise a significant portion of our sample, we have opted not to correct our measured line luminosities for dust attenuation.  

Second, the broad emission lines of many LRDs show signatures of exponential wings that may arise due to scattering effects as opposed to virial motions \citep{Rusakov25, Kokorev26}.  We have decided not to explore the prevalence of exponential profiles in the Skyfire spectra, instead saving that analysis for a future paper.  As a result, our BH masses can be treated as upper limits if there is significant electron scattering or other nonvirial contribution to the line widths.

Third, the single-epoch virial mass relationship of \citet{reines13} was calibrated on relatively local AGN ($z<0.06$) and may no longer be applicable at high redshifts.  That said, \citet{Juodzbalis_2025} recently reported a direct BH mass measurement for an LRD at $z=7.01$ and found it to be in excellent agreement with the mass inferred from single-epoch virial estimates.  This suggests that these relationships may in fact be reliable for LRDs out to relatively high redshifts.

\begin{figure*}[t]
\centering
\includegraphics[width=\linewidth]{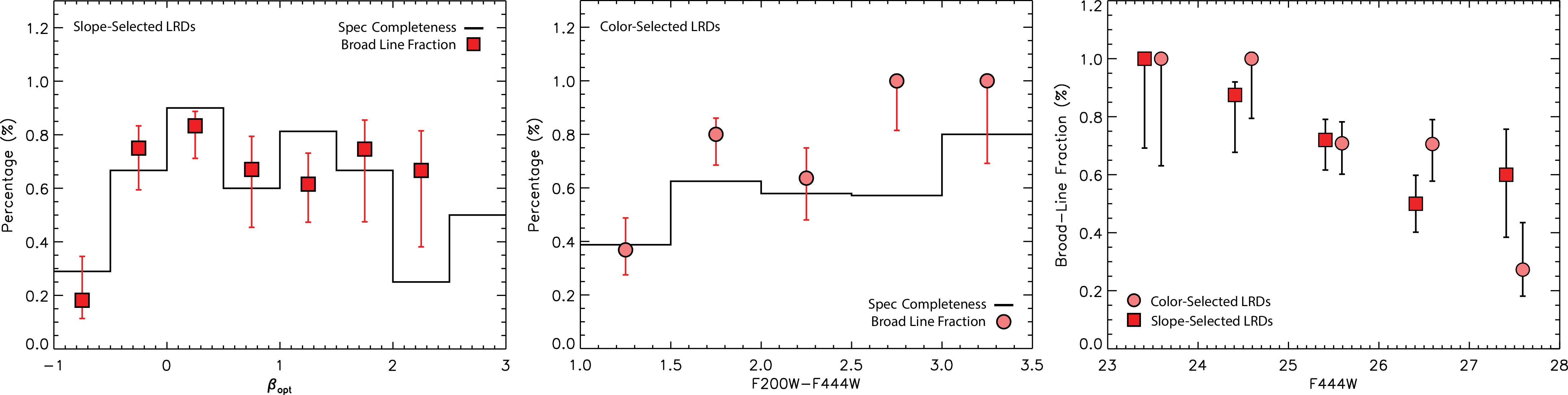}
\caption{(\emph{left}) The spectroscopic completeness and broad-line detection fraction of photometrically-selected LRDs identified using the continuum slope-based criteria of \citet{Kocevski25} versus their rest-optical continuum slope, $\beta_{\rm opt}$.  We find the broad-line fraction remains high below the original optical slope cut of $\beta_{\rm opt} > -0.02$ used in \citet{Kocevski25} and only decreases in the $-1<\beta_{\rm opt} < -0.5$ bin.  (\emph{middle}) The spectroscopic completeness and broad-line fraction of LRDs selected using the color criteria of \citet{Barro26}. We find a decreasing broad-line detection fraction at the bluest F200W-F444W colors.(\emph{right}) The broad-line detection fraction among slope-selected (red) and color-selected (light red) LRDs as a function of their F444W magnitude.  \label{fig:BLfraction}}
\end{figure*}

\section{Continuum Slope \& Color Distribution of Broad-Line LRDs}

With the completion of Skyfire, 73\% of the LRDs from the \citet{Kocevski25} sample in the CEERS field  have been observed spectroscopically with NIRSpec.  In this section, we make use of this relatively high spectroscopic completeness to examine the prevalence of broad emission lines in photometrically-selected LRDs as a function of their observed colors and rest-frame continuum slopes.  

To determine the continuum slope, $\beta$ (where $\beta$ is defined such that $f_{\lambda}\propto\lambda^{\beta})$, for a source, we perform a $\chi^{2}$ minimization fit to the observed magnitudes using the linear relationship: 
\begin{equation}
   m_{i} = -2.5~(\beta+2)~\log {(\lambda_{i})}+c 
\end{equation}
where $m_{i}$ is the AB magnitude measured in the $i$th filter with an effective wavelength of $\lambda_{i}$. We perform this fit to determine both the rest-frame UV and optical spectral slopes, $\beta_{\rm UV}$ and $\beta_{\rm opt}$, using fluxes measured in bands blueward and redward of 3645\AA~based on the redshift of each source as described in \citet{Kocevski25}.  Three bands are used to measure the slope in all cases except for sources at $z>8$, where only the F356W and F444W bands are used to determine the rest-frame optical slope. 

In the left panel of Figure \ref{fig:slope-color}, we show the rest-optical versus rest-UV continuum slopes of the broad line emitters identified with Skyfire.  The red circles indicate LRDs identified in the CEERS imaging using the slope-based selection criteria of \citet{Kocevski25} with their original optical slope cut of $\beta_{\rm opt}>-0.02$ (dotted line) and a lower threshold of $\beta_{\rm opt}>-1$ (dashed line).  LRDs with spectroscopic redshifts are noted as indicated by the legend.  Sources with broad emission lines at $z\gtrsim3$ in the EGS field taken from the literature are also highlighted.  These sources are drawn from \citet{Taylor25}, \citet{Jones26}, \citet{Davis26}, and the recently completed MEGA NIRSpec survey (A.~Kirkpatrick, private communication).   The right panel of Figure \ref{fig:slope-color} shows the F115W-F200W vs F200-F444W color distribution of the Skyfire broad line emitters, along with LRDs identified using the color-based selection criteria of \citet{Barro26}. The dashed lines in that panel denote the LRD selection wedge of \cite{Barro26}.

We can draw several conclusions from the observed slope and color distribution of the LRDs.  First, the broad-line fraction among LRDs remains high below the original optical slope cut of \citet{Kocevski25}.  This can be seen in the left-most panel of Figure \ref{fig:BLfraction}, which shows the spectroscopic completeness and broad-line fraction of slope-selected LRDs versus $\beta_{\rm opt}$.  Of the 18 LRDs with $-0.5 < \beta_{\rm opt} < -0.02$, 12 have been observed spectroscopically with the NIRSpec G395M grating and 9 show signs of broad-line emission ($75.0^{+8.3}_{-15.6}\%$ at 66.7\% spectroscopic completeness).  This is comparable to the broad-line detection fraction in redder LRDs: $73.3^{+5.5}_{-7.5}\%$ for LRDs with $-0.02 < \beta_{\rm opt} < 2.0$ at 78.8\% spectroscopic completeness.

However, at the bluest slopes examined, $-1 < \beta_{\rm opt} < -0.52$, we find that the broad-line fraction drops precipitously to $18.8^{+16.4}_{-6.5}\%$, albeit at a relatively low spectroscopic completeness of 28.9\%.  Of the 38 candidate LRDs with $-1 < \beta_{\rm opt} < -0.52$, 11 have G395M observations and only 2 show broad lines in their spectra.  We conclude that the LRD candidates in this region are likely contaminated by normal galaxies with strong emission lines entering the LRD selection region, resulting in a decreased purity in terms of broad-line recovery.

We find a similar trend of decreasing broad-line detection fraction at the bluest F200W-F444W colors of photometrically-selected LRDs.  This can be seen in the middle panel of Figure \ref{fig:BLfraction}, which shows the spectroscopic completeness and broad-line fraction of color-selected LRDs versus their F200W-F444W color.  At ${\rm F200W-F444W} > 1.5$, $81.4^{+4.5}_{-7.3}\%$ (at 61.4\% spectroscopic completeness) of the color-selected LRD candidates have broad emission lines.  However, at the bluest colors examined, $1.0 < {\rm F200W-F444W} < 1.5$, this fraction drops to $36.8^{+11.8}_{-9.3}\%$ at 38.8\% spectroscopic completeness.  

In addition, we find a low broad-line fraction among LRDs with both red F115W-F200W and F200W-F444W colors, which corresponds to the upper-right spur in the \citet{Barro26} selection wedge.  Of the 7 LRDs near this region that have been observed spectroscopically, two exhibit broad emission lines ($28.6^{+20.4}_{-10.7}\%$ at 58.3\% spectroscopic completeness).  This suggests that among optically red, compact sources, the incidence of broad lines is higher in sources with a stronger UV excess.  In other words, the strength of their v-shaped SEDs.  

The correlation between v-shaped SEDs, compact morphologies and broad emission lines has been noted in the past \citep{Kocevski25, Hviding25} and suggests that a blue rest-UV color is a key criterion to identifying broad-line emitters among this population.  If this UV emission originates from the host galaxy \citep[e.g.,][]{Jones26}, this could indicate that the LRD phenomenon is preferentially associated with systems with young stellar populations and low dust attenuation that intrinsically have blue rest-UV continuum slopes \citep{Topping24}.  Alternatively, it could imply that some component of the UV excess found in LRDs originates from the central engine, possibly in the form of dust-scattered AGN emission \citep{Assef20, Kocevski23b}.  Under this scenario, the observed diversity in UV slopes across the LRD population could reflect varying proportions of emission from the host galaxy and a baseline level of scattered light from the central engine.

\begin{figure}[t]
\centering
\includegraphics[width=3.15in]{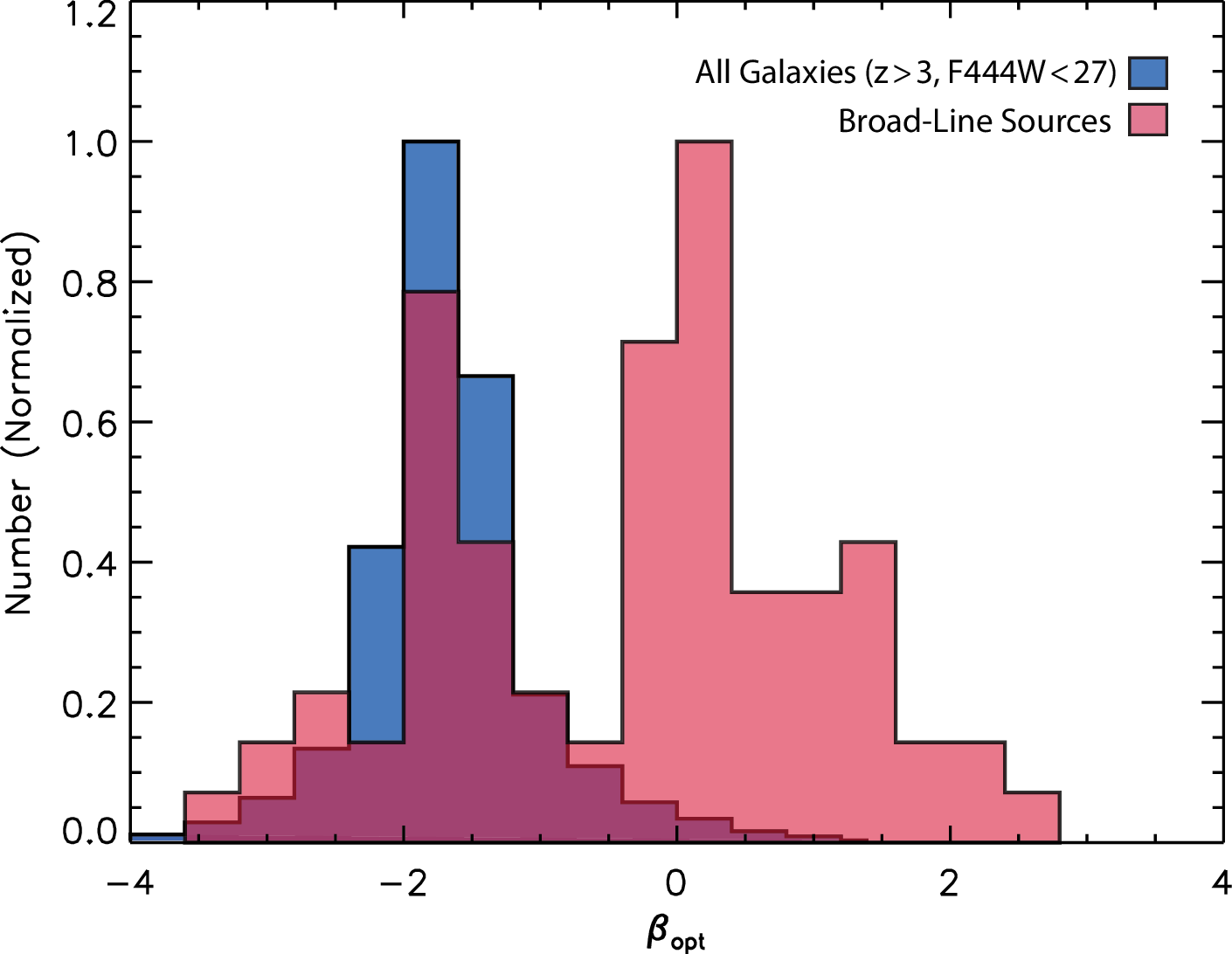}
\caption{The rest-frame optical continuum slope, $\beta_{\rm opt}$, for bright (${\rm F444W}<27$) galaxies (blue histogram) and broad-line emitting sources (red histogram) in the CEERS field at $z>3$.  We find a bimodal distribution between LRDs and the bluest broad-line emitters with a minimum at $\beta_{\rm opt} \sim -1$. \label{fig:beta_opt_distribution}}
\end{figure}

Lastly, we note that the broad-line detection fraction among both slope- and color-selected LRDs is strongly correlated with their F444W magnitude.  This can be seen in the right-most panel of Figure \ref{fig:BLfraction}.  We find that the broad-line fraction among slope-selected LRDs with $\beta_{\rm opt} > -0.52$ drops from an average of $80.9^{+4.6}_{-7.5}\%$ for sources with ${\rm F444W} < 26.5$ to $53.3^{+11.7}_{-12.05}\%$ for sources with ${\rm F444W} > 26.5$.  A commensurate decline is observed in color-selected LRDs.  The H$\alpha$ line luminosity of LRDs is closely correlated to their rest-optical continuum luminosity \citep{Greene24,deGraaff26}, so we attribute this decline with F444W magnitude to the difficulty of finding and confirming broad emission components in sources with weaker emission lines.

\subsection{The Relationship Between LRDs and LBDs}

The connection between LRDs and their bluer counterparts, LBDs, remains a topic of debate.  Recently defined by \citet{Brazzini26} as compact sources with $\beta_{\rm opt} < 0$ and $\beta_{\rm UV} < -0.37$, LBDs share many of the same properties as LRDs, such as their X-ray weakness and a deficiency of hot dust emission, but generally lack Balmer absorption lines indicative of dense gas along the line of sight (\citealt{Jones26,Barro26,Matthee26}; although see \citealt{Scholtz26, Davis26}).  Our finding of a high broad-line detection fraction at bluer rest-optical colors ($\beta_{\rm opt} \sim -0.5$) indicates that the incidence of broad-line AGN extends beyond the reddest colors that define the canonical LRD selection and continues smoothly into the bluer regime occupied by LBDs. This would support the notion that LRDs are the extreme tail of a distribution of broad-line emitters that extends down to at least the reddest LBDs \citep{Billand26} and that these two populations differ primarily in their obscuration level \citep{Madau26} or BH-to-galaxy luminosity ratio \citep{Barro26, Sun26} rather than being two physically distinct classes of astrophysical objects. 

That said, how LRDs and the reddest LBDs relate to the broad-line emitters found with the bluest rest-optical colors ($\beta_{\rm opt} \sim -2$) remains unclear \citep[e.g.,][]{Hainline25}.  When we examine the $\beta_{\rm opt}$ slope distribution of all broad-line emitters in the CEERS field (both LRDs and their blue counterparts), we find a bimodal distribution with relatively few broad-line sources at intermediate slopes ($\beta_{\rm opt}\sim-1$). This distribution can be seen in Figure 7, which shows the $\beta_{\rm opt}$ slopes for bright (${\rm F444W}<27$) galaxies and broad-line emitters detected in the CEERS field at $z>3$. 

\begin{figure}[t]
\centering
\includegraphics[width=3.15in]{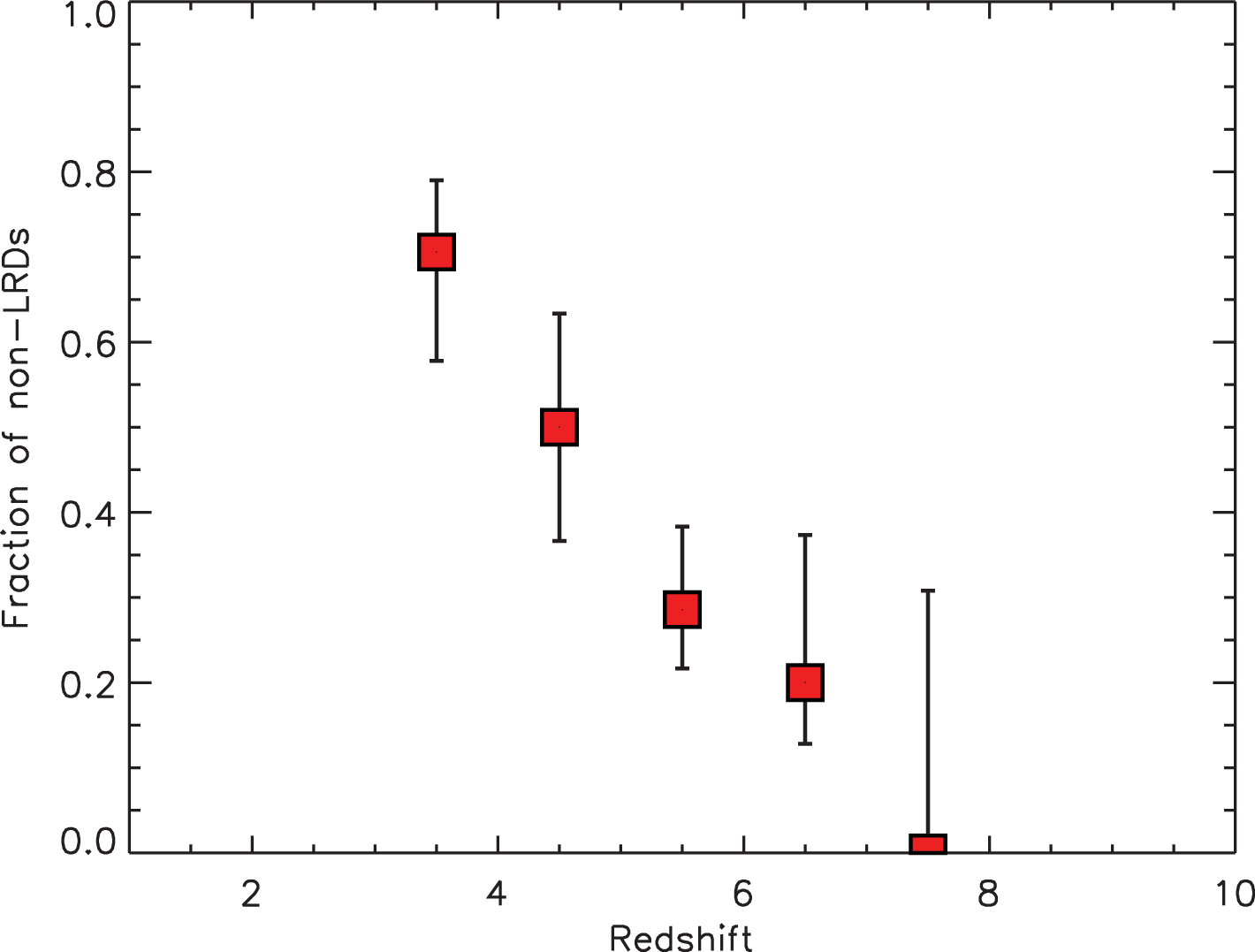}
\caption{The observed fraction of blue broad-line emitting sources (i.e., non-LRDs) in the CEERS field as a function of redshift.  Here blue is defined as $\beta_{\rm opt}<-0.52$.  The fraction increases with decreasing redshift until the blue broad-line sources become the majority of the broad-line population at $z\sim4$. \label{fig:LRDfrac_wz}}
\end{figure}

Taken at face value, this bimodality would seem to argue against a simple unification model where all of the broad-line emitters are members of a single population that differ primarily by their level of obscuration as this would result in a more continuous $\beta_{\rm opt}$ distribution.  However, the observed distribution is undoubtedly affected to some extent by selection effects since LRDs can be directly targeted, while blue broad-line emitters tend to be found serendipitously from among the general galaxy population.  Further directed follow-up of sources with intermediate $\beta_{\rm opt}$ slopes will be needed to determine if the low incidence of broad-line emitters with ``green" rest-optical colors is real.  If this paucity is confirmed, it could indicate that the LRD phase, and its associated reddening, is rapid in its onset and completion and is distinct from the bluest broad-line emitters identified with JWST.  

Recently \citet{Jones26} noted that these blue broad-line sources are preferentially found at lower redshifts ($3<z<4$) and have lower BH-to-stellar mass ratios ($M_{\rm BH}/M_{\star}\sim0.1\%$) compared to canonical LRDs.  This is also borne out by our data and can be seen in Figure \ref{fig:LRDfrac_wz}, which shows the observed fraction of blue broad-line emitters (here defined as $\beta_{\rm opt}<-0.52$, see $\S 9$) in the CEERS field as a function of redshift.  We find the fraction steadily increases with decreasing redshift until the blue broad-line sources become the majority of the population around $z\sim4$.  It is important to note that the observed fraction does not represent the true abundance ratio between LRDs and non-LRDs as this would require correcting for the different selection effects that go into identifying them, a task that is beyond the scope of this paper.  Nonetheless, the evolution of the observed fraction raises the possibility that the bluest broad-line emitters are more evolved versions of their redder counterparts, perhaps found after their dense gas cocoons have cleared and the stellar mass of their host galaxies has built up \citep[e.g.,][]{Inayoshi25}.

\section{Revised LRD Selection Criteria}

In this section, we present a revised version of the LRD selection criteria of \citet{Kocevski25} with changes that are informed by the continuum slope distribution of LRDs with confirmed broad-line detections. 
Given the persistently high broad-line fraction down to $\beta_{\rm opt} \sim -0.5$ and the bimodal $\beta_{\rm opt}$ distribution for broad-line emitters that reaches a minimum at $\beta_{\rm opt} \sim -1$, we propose the following criteria to identify compact objects with v-shaped SEDs and broad emission lines:
\begin{enumerate}[label=(\roman*),leftmargin=2cm]
    \item $\beta_{\rm opt}>-0.52 \,$
    \item $-2.8 < \beta_{\rm UV}<-0.37 \,$
    \item $r_{h} < 1.5~r_{h,~{\rm stars}}.$
\end{enumerate}

\noindent Here $r_{h}$ is the half-light radius of the source measured in the F444W band and $r_{h,~{\rm stars}}$ is the same value measured for stars.  

We find that $80.9^{+4.6}_{-7.5}\%$ of bright sources (${\rm F444W} < 26.5$) that satisfy these criteria in the CEERS field show broad emission lines.  These new cuts effectively incorporate the reddest LBDs into the LRDs classification.  Using this new definition, we find that LRDs make up 54\% of all the faint, broad-line sources identified with JWST in the CEERS field.

The rest-optical slope cut of $\beta_{\rm opt}=-0.52$ corresponds to a F200W-F444W color of 0.75 at $z\sim5$, which is slightly lower than the ${\rm F200W-F444W}>1$ cut proposed by \citet{Barro26}.  Using this lower threshold, we find the F115W-F200W versus F200W-F444W color wedge identifies 93\% of the slope-selected LRDs, with the missing sources found at the extreme ends of the LRD redshift distribution, namely $z<4$ and $z>9$.

\section{Summary \& Conclusions}\label{sec:conclusion}

In this paper, we present the Skyfire program, a 21-hour Cycle 3 JWST/NIRSpec survey with the G395M medium-resolution grating covering five pointings in the CEERS field.  The survey is designed to perform a census of faint, broad-line AGN with a range of rest-optical colors at $z>3$.  The survey's primary targets include photometrically-selected LRDs, blue DLEs and X-ray detected AGN, while filler targets are chosen preferentially from among bright sources in F444W with photometric redshifts of $z>3$.

Skyfire observed 292 sources and recovered secure redshifts for 178, including 34 LRDs, 20 DLE, and 3 X-ray sources.  We find good agreement between our photometric and spectroscopic redshifts, with their difference having a scatter of $\sigma_{\rm NMAD}=0.0348$.  We find the LRDs have a scatter comparable to our filler sample, indicating that the unusual SEDs of LRDs do not result in an increased photometric redshift uncertainty similar to what is commonly found with other types of bright AGN.

In our spectroscopic sample, we identify a total of 34 sources with broad emission lines in the redshift range $2.7 < z < 6.5$.  These lines have FWHMs ranging from 500 to 4000 km s$^{-1}$, the lowest of which likely originate from among the least massive SMBHs known in the early universe.  Two LRDs in our sample have blue-shifted absorption in their Balmer emission lines.  We also find that only 3/17 EELGs selected primarily based on their high-equivalent-width emission lines and compact morphologies exhibit broad emission lines, suggesting that these criteria alone are poorly predictive of broad-line activity. 

With the completion of Skyfire, 73\% of the photometrically-selected LRDs from the \citet{Kocevski25} sample in the CEERS field have been observed spectroscopically.  Combining Skyfire with spectroscopic redshifts and broad-line detections reported in the literature, we explore the prevalence of broad emission lines in LRDs as a function of their rest-frame continuum slope and observed color distributions.  We find the broad-line detection fraction in LRDs remains high at bluer rest-optical colors, suggesting the incidence of broad-line AGN extends beyond the reddest colors that define the canonical LRD selection and extends smoothly into the bluer regime occupied by LBDs. This supports unification models that propose that LRDs and the reddest LBDs differ primarily in their obscuration level or BH-to-galaxy luminosity ratio rather than being two physically distinct classes of astrophysical objects. 

We do, however, find a bimodal $\beta_{\rm opt}$ slope distribution between LRDs and broad-line sources with the bluest rest-optical colors.  We show that the observed fraction of blue ($\beta_{\rm opt}<-0.52$) broad-line emitting sources increases with decreasing redshift until they become the majority of the broad-line population at $z\sim4$.  This raises the possibility that the bluest broad-line emitters may be more evolved versions of their redder counterparts, found after their dense gas cocoons have cleared out and the stellar mass of their host galaxies has built up.

We also find a low broad-line fraction in photometrically-selected LRD with the reddest rest-UV colors, suggesting the strength of the v-shape in the SEDs of LRDs is closely tied to the incidence of broad-line emission.  This could imply that some component of the UV excess found in LRDs originates from the central engine and that the diversity in UV slopes across the LRD population is due to varying proportions of light from the host galaxy and a baseline level of emission from the central engine, possibly in the form of dust-scattered light. 

Finally, we present revised LRD selection criteria for identifying compact objects with v-shaped SEDs and broad emission lines that extends down to $\beta_{\rm opt}=-0.52$.  Using these criteria, we find that $80.9^{+4.6}_{-7.5}\%$ of bright (${\rm F444W} < 26.5$), photometrically-selected LRDs in the CEERS field have broad emission lines.  With this lower $\beta_{\rm opt}$ threshold, we find that LRDs make up 54\% of all the broad-line emitters found with JWST in the CEERS field.

In the future, we aim to use Skyfire to examine the spectra of individual sources of interest, as well as continue our study of LRD demographics, which is possible due to the high spectroscopic completeness achieved in the CEERS field.  The data collected by Skyfire is publicly available and we welcome and encourage its use by the broader community.

\section{Acknowledgments}

The Skyfire team would like to thank our program coordinators, Alaina Henry and Weston Eck, for their help with planning and executing our program.  We also thank the previous surveys that contributed to making the EGS such a data rich field, including AEGIS, CANDELS, 3D-HST, CEERS, RUBIES, CAPERS, THRILS, C3PO, OCEANS, and MEGA. This work is supported by NASA grant JWST-GO-5718 based on observations made with the NASA/ESA/CSA James Webb Space Telescope. The data were obtained from the Mikulski Archive for Space Telescopes at the Space Telescope Science Institute, which is operated by the Association of Universities for Research in Astronomy,
Inc., under NASA contract NAS 5-03127 for JWST.

\appendix

\section{NIRSpec Spectra of Broad Line Sources}

\begin{figure*}[t]
\centering
\includegraphics[width=\linewidth]{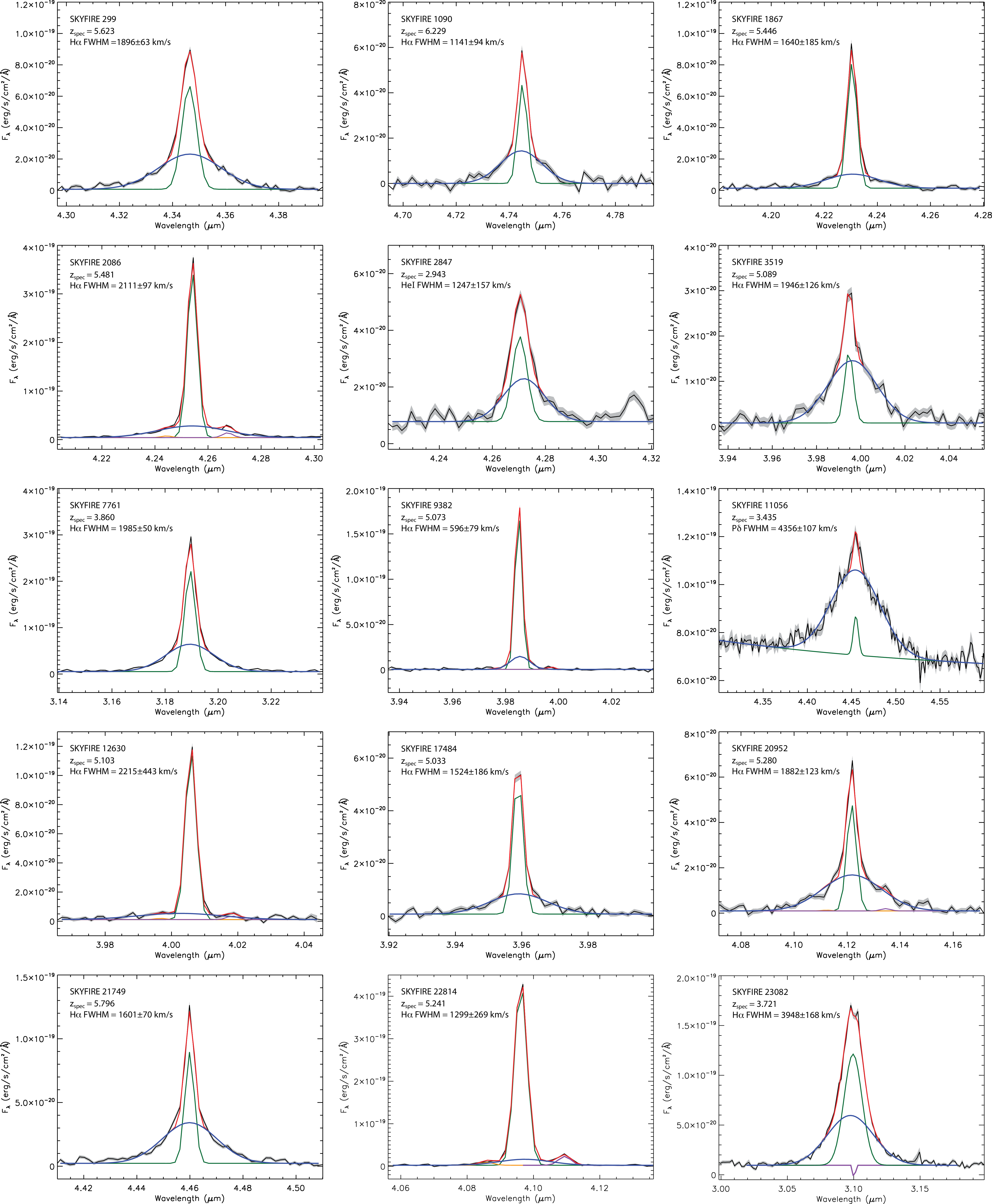}
\caption{NIRSpec spectra with uncertainties (grey shaded region) taken in the G395M grating of Skyfire sources that feature broad emission lines detected at a ${\rm SNR}>5$. Green lines show the best-fit Gaussian for the narrow emission line component, blue lines show the best-fit broad component, and red lines show the best overall (narrow plus broad) fit the emission line.  The FWHM of the broad component (corrected for instrument broadening) is shown in the upper left of each panel. \label{fig:BLfits1}}
\end{figure*}

Here we present the G395M spectra of the 33 new Skyfire sources that show broad emission features, in regions near the broad component.  These spectra, along with our best-fit two component (narrow plus broad) emission line model for each line, are shown in Figure \ref{fig:BLfits1}.

\renewcommand{\thefigure}{\arabic{figure} (Cont.)}
\addtocounter{figure}{-1}
\begin{figure*}[t]
\centering
\includegraphics[width=\linewidth]{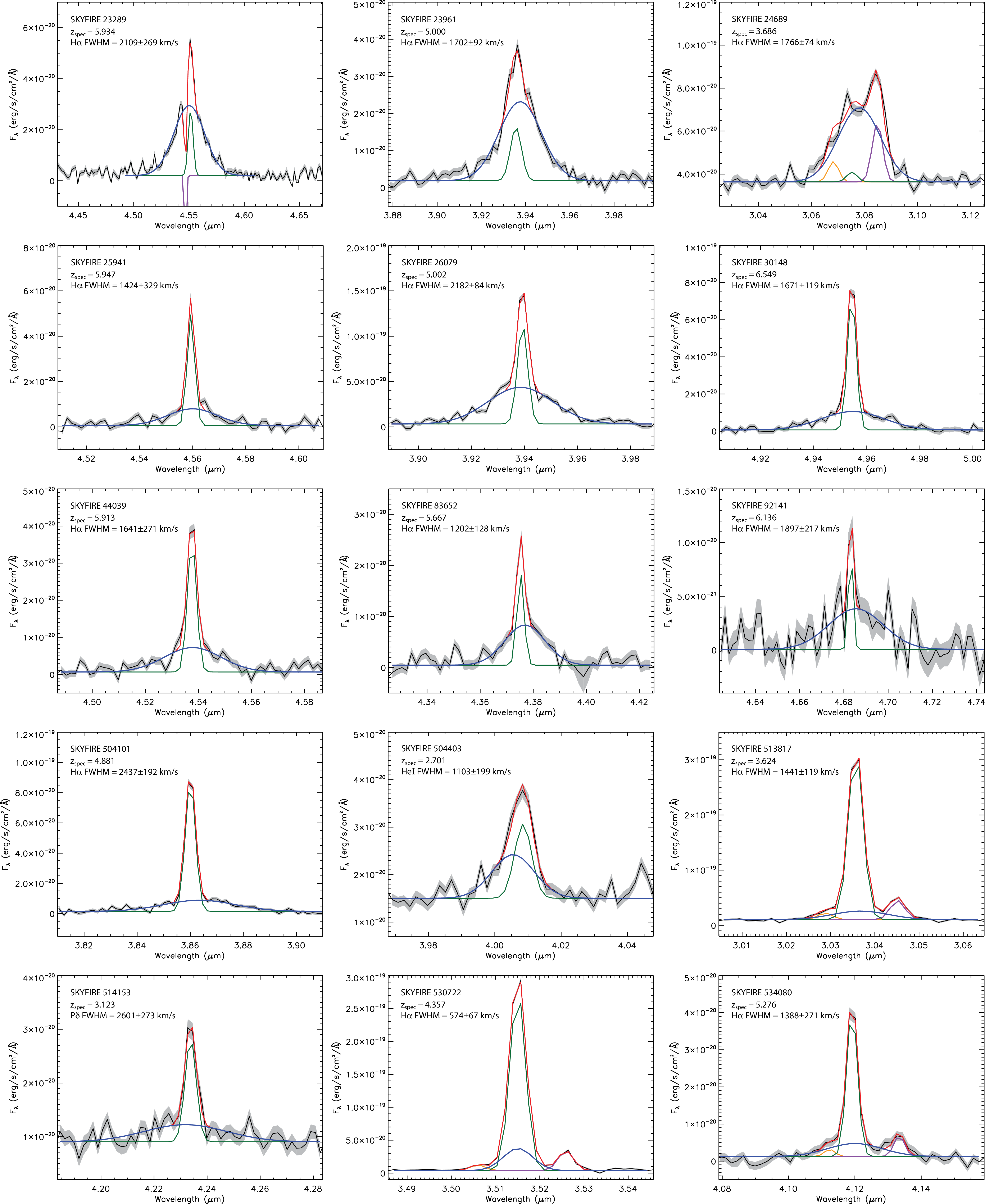}
\caption{NIRSpec spectra with uncertainties (grey shaded region) taken in the G395M grating of Skyfire sources that feature broad emission lines detected at a ${\rm SNR}>5$. Green lines show the best-fit Gaussian for the narrow emission line component, blue lines show the best-fit broad component, and red lines show the best overall (narrow plus broad) fit the emission line.  The FWHM of the broad component (corrected for instrument broadening) is shown in the upper left of each panel. \label{fig:BLfits2}}
\end{figure*}

\renewcommand{\thefigure}{\arabic{figure} (Cont.)}
\addtocounter{figure}{-1}

\begin{figure*}[t]
\centering
\includegraphics[width=\linewidth]{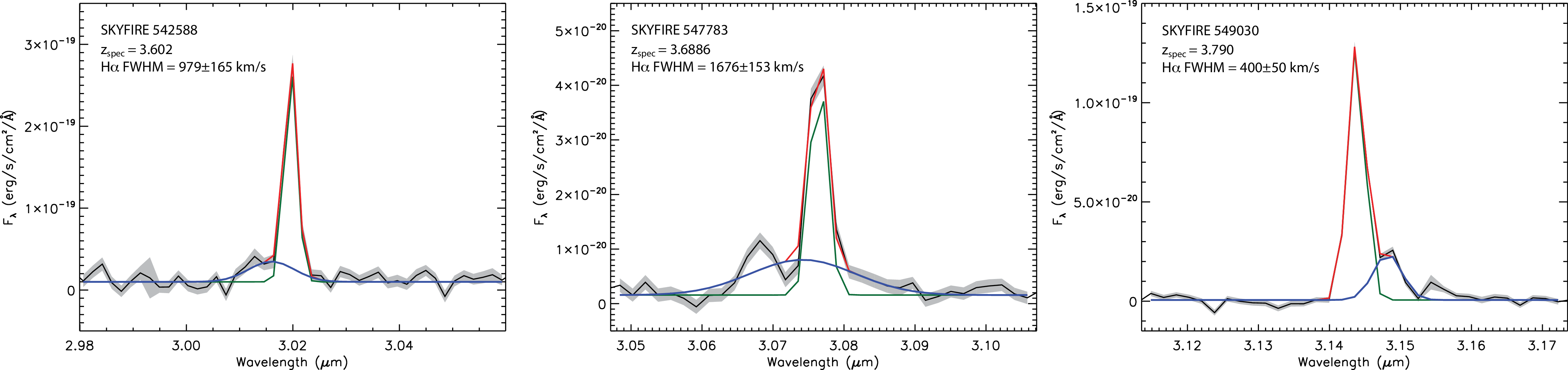}
\caption{NIRSpec spectra with uncertainties (grey shaded region) taken in the G395M grating of Skyfire sources that feature broad emission lines detected at a ${\rm SNR}>5$. Green lines show the best-fit Gaussian for the narrow emission line component, blue lines show the best-fit broad component, and red lines show the best overall (narrow plus broad) fit the emission line.  The FWHM of the broad component (corrected for instrument broadening) is shown in the upper left of each panel. \label{fig:BLfits3}}
\end{figure*}

\renewcommand{\thefigure}{\arabic{figure}}

\bibliography{refs}{}


\end{document}